%% file: String_lattice_8.tex
\documentclass[11pt]{article}
\usepackage{booktabs}
\usepackage{pgfplotstable}
\usepackage{float}
\usepackage{amssymb,amsmath,amsfonts, mathtools}
\usepackage{enumerate}
\usepackage{dsfont}
\usepackage[sort,compress]{cite}
\usepackage{verbatim}
\usepackage{amsfonts,euscript,amssymb,stmaryrd,braket}
\usepackage{graphics,tikz}
\usetikzlibrary{arrows,decorations.markings,patterns}
\usepackage{caption}
\usepackage{subcaption}
\usepackage{slashed}
\definecolor{hyperref}{RGB}{026,028,185}
\usepackage[bookmarks=true,colorlinks=true,linkcolor=hyperref,citecolor=hyperref,urlcolor=hyperref,bookmarksnumbered]{hyperref}
 \usepackage{booktabs}
 \usepackage{multirow}
\usepackage{tabularx}
\usepackage{longtable}
\usepackage{bbold,bbding}
\usepackage{amsthm} 

 \def\clock{{\count0=\time
           \divide\count0 60
           \ifnum\count0<10 0\fi\the\count0
           \multiply\count0 -60 \advance\count0 \time
           :\ifnum\count0<10 0\fi \the\count0
         }}
\newcommand{\timestamp}{{\small\vbox{\hbox{\tt\jobname.tex}
\hbox{\the\day/\the\month/\the\year, \clock}}}}

\newcommand{\ba}{\begin{eqnarray}}
\newcommand{\ea}{\end{eqnarray}}

\newcommand{\be}{\begin{equation}}
\newcommand{\ee}{\end{equation}}

\makeatletter
\let\old@startsection=\@startsection
\let\oldl@section=\l@section
\renewcommand{\@startsection}[6]{\old@startsection{#1}{#2}{#3}{#4}{#5}{#6\mathversion{bold}}}
\renewcommand{\l@section}[2]{\oldl@section{\mathversion{bold}#1}{#2}}
\makeatother

\numberwithin{equation}{section}
\usepackage{color}

\def\la{\label}
\def\no{\nonumber}
\def\a{\alpha}
\def\b{\beta}

\def \adss {$AdS_5 \times S^5$\ }

\begin{document}
\renewcommand{\thefootnote}{\arabic{footnote}}

\overfullrule=0pt
\parskip=2pt
\parindent=12pt
\headheight=0in \headsep=0in \topmargin=0in \oddsidemargin=0in

\vspace{ -3cm} \thispagestyle{empty} \vspace{-1cm}
\begin{flushright} 
\footnotesize
HU-EP-16/11\\
QMUL-PH-16-09
\end{flushright}%

\begin{center}
\vspace{1.2cm}
{\Large\bf \mathversion{bold}
Green-Schwarz superstring on the lattice
}
 
\author{ABC\thanks{XYZ} \and DEF\thanks{UVW} \and GHI\thanks{XYZ}}
 \vspace{0.8cm} {
 L.~Bianchi$^{a,b,}$\footnote{ {\tt  lorenzo.bianchi@desy.de}}, M.~S.~Bianchi$^{c,}$\footnote{ {\tt m.s.bianchi@qmul.ac.uk}},  V.~Forini$^{a,}$\footnote{{\tt $\{$valentina.forini,leder,edoardo.vescovi$\}$@\,physik.hu-berlin.de}}, B.~Leder$^{a,3}$,  E.~Vescovi$^{a,3}$}
 \vskip  0.5cm

\small
{\em
$^{a}$Institut f\"ur Physik, Humboldt-Universit\"at zu Berlin, IRIS Adlershof, \\Zum Gro\ss en Windkanal 6, 12489 Berlin, Germany  
\vskip 0.05cm
$^{b}$        II. Institut f\"ur Theoretische Physik, Universit\"at Hamburg,\\ Luruper Chaussee 149, 22761 Hamburg, Germany \vskip 0.05cm
$^{c}$ Queen Mary University of London, Mile End Road, London E1 4NS, UK}
\normalsize

\end{center}

\vspace{0.3cm}
\begin{abstract} 
We consider possible discretizations for a gauge-fixed Green-Schwarz action of Type IIB superstring. We use them for measuring the action, from which we extract the cusp anomalous dimension of planar $\mathcal{N}=4$ SYM as derived from  AdS/CFT, as well as the mass of the two $AdS$ excitations  transverse to the relevant null cusp classical string solution.         
We perform lattice simulations employing  a Rational Hybrid Monte Carlo (RHMC) algorithm and two Wilson-like fermion discretizations, one of which preserves the global $SO(6)$ symmetry of the model.
We compare our results with the expected behavior at various values of  $g=\frac{\sqrt{\lambda}}{4\pi}$. 
For both the observables, we find a good agreement  for large $g$, which is the perturbative regime of the sigma-model. For smaller values of $g$,  
the expectation value of the action exhibits a deviation compatible with the presence of quadratic divergences. 
After their non-perturbative subtraction the continuum limit can be taken, and suggests a qualitative agreement with the non-perturbative expectation from AdS/CFT. 
Furthermore, we  detect a  phase in the fermion determinant, whose origin we explain, that  for small $g$ leads  to a sign problem not treatable via standard reweigthing.   The continuum extrapolations of the  observables in the two different discretizations agree within errors, which is strongly suggesting that they lead to the same continuum limit. \\ Part of the results discussed  here  were presented earlier in~\cite{POS2015}.
\end{abstract}

\newpage

%%%%%%%%%%%%%%%%%%%%%%%%%%%%%%%%%%%%%%%%%%%%%
%%%%%%%%%%%%%%%%%%%%%%%%%%%%%%%%%%%%%%%%%%%%%
\tableofcontents
  
\section{Introduction}

The maximally supersymmetric and superconformal $\mathcal{N}=4$ super Yang-Mills (SYM) theory is a unique example of non-trivially interacting, four-dimensional gauge theory which is believed to be exactly integrable~\cite{Beisert_review}.  
A plethora of results, obtained relying on the assumption of an all-loop integrability for this model and exploiting therefore sophisticated Bethe-Ansatz-like techniques,  have been confirmed by direct  perturbative  computations both in gauge theory and in its AdS/CFT dual - the Type IIB, Green-Schwarz string  propagating in the maximally supersymmetric background $AdS_5\times S^5$ supported by a self-dual Ramond-Ramond (RR) five-form flux. 
%The ``might be'' above refers to the tests of integrability-related predictions in the \emph{finite coupling} region, which is as such not the realm of  perturbation theory on both sides of the AdS/CFT system and where supersymmetric localization can be a powerful computational tool only for a restricted class of BPS-protected observables. 
%Without the assumption of quantum integrability, the model can be solved at finite coupling via supersymmetric localization techniques~\cite{Pestun}. These, however, apply only to a restricted class of BPS-protected observables and are only defined in the field theory.  
Without the assumption of quantum integrability, a restricted class of BPS-protected observables can be computed at finite coupling via supersymmetric localization techniques~\cite{Pestun}, which are however only defined on the field theory side.
The superstring sigma-model,  for which integrability is a solid fact only classically,  is a complicated, highly non-interacting 2d theory which is under control only perturbatively~\footnote{See~\cite{Tseytlin:2010jv,McLoughlin:2010jw} for reviews,~\cite{Roiban:2007jf,Roiban:2007ju,Giombi, Giombi:2010bj, Giombi:2010zi, abjm2loops, 
Bianchi:2015laa,Bianchi:2015iza,Bianchi:2015vgw} for studies of the models of interest here and~\cite{Gubser:2002tv,Frolov:2002av, Frolov:2003qc, McLoughlin:2008he, Beccaria:2008tg, Beccaria:2010ry, Drukker:2011za, Forini:2010ek, Forini:2012bb, Bianchi:2013nra, Forini:2014kza, Forini:2015mca, Forini:2015bgo} for related studies.}.

The natural, genuinely field-theoretical way to investigate the finite-coupling region and in general the non-perturbative realm of a quantum field theory  is to discretize the spacetime where the model lives, and proceed with numerical methods for the  lattice field theory so defined.  A rich and interesting program of putting $\mathcal{N}=4$ SYM on the lattice is being carried out for some years by Catterall et al.~\cite{Catterall_physrept,Bergner:2016sbv,Schaich:2015ppr} (see also~\cite{Hanada:2016jok} for a report on further uses of lattice techniques in problems relevant in AdS/CFT). 
Alternatively, one could discretize the worldsheet spanned by the Green-Schwarz string embedded in $AdS_5\times S^5$.  
%, and study for example -- as we will do here -- the corresponding partition function, with the appropriate boundary conditions to be representing a certain Wilson loop in the dual gauge theory. 
This much less explored 
route has been first proposed in~\cite{Roiban}, where the  most studied observable of the AdS/CFT integrable system   - the  cusp anomaly of $\mathcal{N}=4$ SYM -  has been investigated with lattice techniques from the point of view of string theory. 

In this paper, we revisit and extend the analysis of~\cite{Roiban}. We will therefore discretize the two-dimensional worldsheet spanned classically by an open string ending, at the $AdS_5$ boundary where the four-dimensional field theory lives, on a light-like cusp which is the countour of the dual Wilson loop.  The renormalization of the latter is governed by the cusp anomaly $f(g)$, a function of the coupling  $g=\frac{\sqrt{\lambda}}{4\pi}$  ($\lambda$ is the 't Hooft coupling of the AdS/CFT dual gauge theory)~\footnote{In the AdS/CFT context, where the 't Hooft coupling $\lambda\sim g^2$ is used as relevant parameter, the large $g$ region is naturally referred to as ``strong coupling'' regime. 
The string worldsheet sigma-model of interest here, for which perturbation theory is a  $1/g$ expansion,  is however weakly-coupled at large $g$.}
which in this framework is often simply referred to as ``scaling function''~\footnote{The  ``scaling function'' $f(g)$ is in fact the coefficient of $\log S$  in the large spin $S$  anomalous dimension $\Delta$  of leading twist operators $\Delta=f(g)\log S+{\cal O}(\log S/S)$. It  equals~\cite{Korchemsky:1992xv}  twice 
the cusp anomalous dimension $\Gamma_{\rm cusp}$ of light-like 
Wilson loops%, in terms of which \eqref{cusp_WL} is actually written
\begin{equation}\label{gammacusp}
\langle W[C_{\rm cusp}]\rangle \sim e^{-\Gamma_{\rm cusp}\,\gamma\, \ln \frac{\Lambda_{UV}}{\Lambda_{IR}} }\,,
\end{equation}
where $\gamma$ is the large, real parameter related to the geometric angle $\phi$ of the cusped Wilson loop
by $i\gamma=\phi$. The expectation value above is in fact extracted in the large imaginary $\phi$ limit.
The same function $f(g)$ also governs the  infrared structure of gluon scattering amplitudes. }. 
%\begin{equation}\label{cusp_WL}
%\langle W[C_{\rm cusp}]\rangle \sim e^{-\frac{1}{2} f(g)\,\gamma\, \ln \frac{\Lambda_{UV}}{\Lambda_{IR}} }\,,\qquad 
%\end{equation}
According to  AdS/CFT, any Wilson loop expectation  value %(\ref{cusp_WL}) 
should be represented by the  path integral of an open string ending at the AdS boundary~\cite{Maldacena:1998im,Rey:1998ik}, in this case 
\begin{equation}\label{Z_cusp}
 \langle W[C_{\rm cusp}]\rangle\equiv  Z_{\rm cusp}= \int [D\delta X] [D\delta\Psi]\, e^{- S_{\rm cusp}[X_{\rm cl}+\delta X,\delta\Psi]} = e^{-\Gamma_{\rm eff}}\equiv e^{-\frac{1}{8} f(g)\,V_2 }~.                         
\end{equation}
Above, $X_{\rm cl}=X_{\rm cl}(t,s)$ - with $t,s$ the temporal and spatial coordinate spanning the string worldsheet -  is the classical solution of the string equations of motion describing the world surface of an open string ending on a null cusp\cite{Giombi}. This vacuum, also known as GKP~\cite{Gubser:2002tv} string, is of crucial and persisting importance in AdS/CFT, as  holographic dual to several fundamental observables in the gauge theory~\cite{Alday:2007mf}
which can be studied exploting the underlying integrability of the AdS/CFT system (see e.g.~\cite{Alday:2010ku, Basso:2013vsa,Basso:2013aha}). $S_{\rm cusp}[X+\delta X,\delta\Psi]$ is the action for field fluctuations over it -- the fields being both bosonic and fermionic string coordinates $X(t,s),~\Psi(t,s)$ -- and is reported below in equation \eqref{S_cusp} in terms of the effective bosonic and fermionic degrees of freedom remaining after gauge-fixing. Since the fluctuation Lagrangian has constant coefficients, the worldsheet volume $V_2=\int dt ds$  simply factorizes out~\footnote{As mentioned above, $f(g)$ equals twice the coefficient of the logarithmic divergence in \eqref{gammacusp}, for which the stringy counterpart should be the infinite two-dimensional worldsheet volume. The further normalization of $V_2$ with a $1/4$ factor follows the convention of~\cite{Giombi}.} in front of the function of the coupling  $f(g)$, as in the last equivalence in \eqref{Z_cusp}.  
The scaling function $f(g)$ can be evaluated perturbatively in gauge theory~\cite{Bern:2006ew} ($g\ll1$), and  in  sigma-model loop expansion~\cite{Gubser:2002tv,Frolov:2002av,Giombi} ($g\gg 1$) as in \eqref{cuspperturbative} below.  
%\begin{flalign}\la{cuspperturbative}
%f(g)=\begin{cases} 
%8g^2\Big[1-\frac{\pi^2}{3}g^2+\frac{11\,\pi^2}{45} g^2-\Big(\frac{73}{315}+8\,\zeta_3\Big)g^6+\cdots\Big]& 
%\,,\qquad\qquad g\ll1\\ 
%4\,g\,\Big[1-\frac{3\ln2}{4\pi\,g}-\frac{K}{16\,\pi^2\,g^2}+\cdots\Big]& 
%\,,\qquad\qquad g\gg1\,.
%\end{cases}
%\end{flalign}
Assuming all-order integrability of the spectral problem for the relevant operators and taking a thermodynamic  limit of the corresponding asymptotic Bethe Ansatz, an integral equation~\cite{BES} can be derived which gives $f(g)$ exactly at each value of the coupling, and when  expanded in the corresponding regimes gives back \eqref{cuspperturbative}.

Rather than partition functions, in a lattice approach it is natural to study  vacuum expectation values. In simulating  the vacuum expectation value of the ``cusp'' action  
\begin{eqnarray}\label{vevaction}
%  S&=&g\,a^2\sum \mathcal{L}\\
\langle S_{\rm cusp}\rangle&=& \frac{ \int [D\delta X] [D\delta\Psi]\, S_{\rm cusp}\,e^{- S_{\rm cusp}}}{ \int [D\delta X] [D\delta\Psi]\, e^{- S_{\rm cusp}}} = -g\,\frac{d\ln Z_{\rm cusp}}{dg}\equiv g\,\frac{V_2}{8}\,f'(g)   ~,
  \end{eqnarray}
  we are therefore supposed to obtain information on the \emph{derivative} of the scaling function~\footnote{Here our analysis is different from the one in~\cite{Roiban}.  In particular,  $\langle S\rangle\sim\frac{f(g)}{V_2/2}$ only when $f(g)$ is  linear in $g$, which happens as from \eqref{cuspperturbative}   for \emph{large} $g$. }.  

\bigskip

It is important to emphasize that the analysis here carried out is far from being a non-perturbative definition, \`a la Wilson lattice-QCD, of the Green-Schwarz  worldsheet string model. For this purpose one should work with a Lagrangian which is invariant under the local symmetries - bosonic diffeomorphisms and $\kappa$-symmetry -  of the model,  while below we will make use of an  action which \emph{fixes them all}. There is however a number of reasons which make this model  interesting for lattice investigations, within and hopefully beyond the community interested in holographic models. 
If the aim is a test of holography and integrability, it is obviously computationally cheaper to use a two-dimensional grid, rather than a four-dimensional one, where no gauge degrees of freedom are present and all fields are assigned to sites - indeed, only scalar fields (some of which anticommuting) appear in $S_{\rm cusp}$. Also, although we are dealing with superstrings, there is here no subtlety involved with putting supersymmetry on the lattice, both because of the Green-Schwarz formulation of the action (with supersymmetry only manifest in the target space)~\footnote{In perturbation theory, both in the continuum and on the lattice, one can however observe an effective two-dimensional supersymmetry spontaneously broken by the classical solution. This ensures a non-vanishing, finite (due to mass - squared sum rule) vacuum energy.} and because $\kappa$-symmetry is gauge-fixed.  As computational playground this is an  interesting one on its own, allowing in principle for explicit investigations/improvements of algorithms: a highly-nontrivial two-dimensional model with four-fermion interactions, for which  relevant observables have not only, through AdS/CFT,  an explicit analytic  strong coupling expansion -- the perturbative series in the dual gauge theory  -- but also, through AdS/CFT \emph{and} the assumption of integrability, an explicit numerical prediction at all couplings. 
In general, one merit of the analysis initiated in~\cite{Roiban} and that we readdress here is to explore another route via which  lattice simulations could become a potentially  efficient tool in numerical holography (see also~\cite{Roiban} for a discussion on further examples of interesting observables that could be investigated this way).  
%
%Clearly, a successful simulation 
%In this sense - while one could state that we completely bypass the subtleties of $\kappa$-symmetry 

\bigskip

The paper proceeds as follows. In Section \ref{sec:continuum} we describe in the continuum the model and the linearization of its quartic fermionic interactions~\cite{Roiban}. 
In Section \ref{sec:discretization} we present the $SO(6)$-preserving Wilson-like discretization adopted for the simulations shown in the main body. 
In Section \ref{sec:simulation}, after commenting on the way we perform the continuum limit,  we show the result of our measurements for  the correlator of two bosonic fields (the $AdS$ lagrangean excitations  transverse to the classical string solution), for the expectation value of the action \eqref{vevaction}, and for a complex phase implicit in the linearization.
Conclusions are drawn in Section \ref{sec:conclusions}. Details of the model in the continuum and on an  alternative discretization used are collected in  Appendices \ref{app:continuum} and \ref{app:altern_discretization} respectively.

 %%%%%%%%%%%%%%%%%%%%%%
 \section{The model in the continuum and its linearization}
 \label{sec:continuum}
%%%%%%%%%%%%%%%%%%%%%%

In the continuum, the   \adss   superstring  ``cusp'' action, which describes quantum  fluctuations above the null cusp background  can be written after Wick-rotation as~\cite{Giombi}
\begin{eqnarray}\nonumber
&& S_{\rm cusp}=g \int dt ds \mathcal{L}_{\rm cusp}\\\nonumber
&& \!\!\!\!\!\! 
\mathcal{L}_{\rm cusp} = |\partial_{t}x+\!\textstyle{\frac{1}{2}}x|^{2}+\!\frac{1}{ {z}^{4}} |\partial_{s} {x}-\!\!\textstyle{\frac{1}{2}} {x}|^{2}+\left(\partial_{t}z^{M}+\frac{1}{2} {z}^{M}+\!\frac{i}{ {z}^{2}} {z}_{N} {\eta}_{i}\left(\rho^{MN}\right)_{\phantom{i}j}^{i} {\eta}^{j}\right)^{2}+\frac{1}{ {z}^{4}}\left(\partial_{s} {z}^{M}-\textstyle{\frac{1}{2}} {z}^{M}\right)^{2}  \\\label{S_cusp}
 && \!\!\!\!\!\!
  +i\left( {\theta}^{i}\partial_{t}{\theta}_{i}+ {\eta}^{i}\partial_{t}{\eta}_{i}+ {\theta}_{i}\partial_{t}{\theta}^{i}+ e{\eta}_{i}\partial_{t} {\eta}^{i}\right)-\textstyle{\frac{1}{{z}^{2}}}\left( {\eta}^{i}{\eta}_{i}\right)^{2}  \\\nonumber
 &&  \!\!\!\!\!\!
 +2i\Big[\textstyle\frac{1}{z^{3}}z^{M} {\eta}^{i}\left(\rho^{M}\right)_{ij}
 \left(\partial_{s} \theta^j-\!\frac{1}{2} \theta^j-\!\!\frac{i}{{z}} {\eta}^{j}\left(\partial_{s} {x}-\!\!\frac{1}{2} {x}\right)\right)\!+\!\frac{1}{{z}^{3}}{z}^{M}{\eta}_{i} (\rho_{M}^{\dagger} )^{ij}\left(\partial_{s}{\theta}_{j}-\!\frac{1}{2}{\theta}_{j}+\!\frac{i}{{z}}{\eta}_{j}\left(\partial_{s}{x}-\frac{1}{2}{x}\right)\!^{*}\right)\!\!\Big]
\end{eqnarray}
Above, $x,x^*$ are the two bosonic $AdS_5$ (coordinate) fields transverse  to the $AdS_3$ subspace of the classical solution. Together with $z^M\, (M=1,\cdots, 6)$ ($z=\sqrt{z_M z^M}$), they are the bosonic coordinates of the $AdS_5\times S^5$ background in Poincar\'e parametrization remaining after fixing a ``AdS light-cone gauge''~\cite{MT2000,MTT2000}. 
In Appendix \ref{app:continuum} we briefly  review the steps leading to the action \eqref{S_cusp}.  
 %As mentioned above,  the lagrangean above neither contains gauge fields nor actual fermions. Indeed, 
 The fields $\theta_i,\eta_i,\, i=1,2,3,4$ are  4+4 complex anticommuting variables for which  $\theta^i = (\theta_i)^\dagger,$ $\eta^i = (\eta_i)^\dagger$. They transform in the fundamental representation of the $SU(4)$ R-symmetry and do not carry (Lorentz) spinor indices.  The matrices $\rho^{M}_{ij} $ are the off-diagonal
blocks of $SO(6)$ Dirac matrices $\gamma^M$ in the chiral representation
\begin{equation} 
\gamma^M\equiv \begin{pmatrix}
0  & \rho^\dagger_M   \\
 \rho^M   &  0  
\end{pmatrix}
=
\begin{pmatrix}
0  & (\rho^M)^{ij}   \\
(\rho^M)_{ij}   &  0 
\end{pmatrix}
\end{equation}
The two off-diagonal blocks, carrying upper and lower indices respectively, are related by $(\rho^M)^{ij}=-(\rho^M_{ij})^*\equiv(\rho^M_{ji})^*$, so that indeed the block with upper indices, denoted  $(\rho_{M}^{\dagger})^{ij}$, is the conjugate transpose of the block with lower indices.
$(\rho^{MN})_i^{\hphantom{i} j} = (\rho^{[M} \rho^{\dagger N]})_i^{\hphantom{i} j}$ and
$(\rho^{MN})^i_{\hphantom{i} j} = ( \rho^{\dagger [M} \rho^{N]})^i_{\hphantom{i} j}$ are  the
$SO(6)$ generators. 
%The fields $z^M$ are neutral under U(1), $\theta^i$ and $\eta^i$ have opposite
%charges and the charge of $\eta_i$ is half the charge of $x$. 

In the action \eqref{S_cusp}, as standard in the literature,  the light-cone momentum has been consistently set to the unitary value, $p^+=1$. Clearly, in the perspective adopted here it is crucial to keep track of dimensionful quantities, which are in principle subject to renormalization. In the following we will make explicit the presence of one massive  parameter,  defined as $m$, as well as its dimensionless counterpart $M=a\,m$. The latter and the (dimensionless) $g$ are the only ``bare''
 parameters characterizing  the model in the continuum. %, where dimensionless quantities -- as $f(g)$ -- will depend only on $g$.
 
In \eqref{S_cusp},  local bosonic (diffeomorphism) and fermionic ($\kappa$-) symmetries originally present in the Type IIB superstring action on $AdS_5\times S^5$~\cite{MT1998} have been fixed in a ``AdS light-cone gauge''~\cite{MT2000,MTT2000}. On the other hand two important global symmetries are explicitly realized. The first one is the $SU(4)\sim SO(6)$ symmetry originating from the isometries of $S^5$, which is unaffected by the gauge fixing. Under this symmetry the fields $z^M$ change in the $\mathbf{6}$ representation (vector representation), the fermions $\{\eta_i,\theta_i\}$ and $\{\eta^i,\theta^i\}$ transform in the $\mathbf{4}$ and $ \mathbf{\bar 4}$ (fundamental and anti-fundamental) respectively, whereas the fields $x$ and $x^*$ are simply neutral. The second global symmetry is a $SO(2)\sim U(1)$ arising from the rotational symmetry in the two $AdS_5$ directions orthogonal to $AdS_3$ (i.e. transverse to the classical solution) and therefore, contrary to the previous case, the fields $x$ and $x^*$ are charged (with charges $1$ and $-1$ respectively) while the $z^M$ are neutral. The invariance of the action simply requires the fermions $\eta_i$ and $\theta^i$ to have charge $\frac12$ and consequently $\eta^i$ and $\theta_i$ acquire charge $-\frac12$. An optimal discretization should preserve  the full global  symmetry of the model. In Section \ref{sec:discretization} we will see that in the case of the $SO(2)$ symmetry this is not possible.  

With the action \eqref{S_cusp} one can directly proceed to the perturbative evaluation of the effective action in~\eqref{Z_cusp}, as done  in~\cite{Giombi} up to two loops in sigma-model perturbation theory, obtaining for the cusp anomaly  ($K$ is the Catalan constant)
\begin{flalign}\la{cuspperturbative}
f(g)=4\,g\,\Big(1-\frac{3\log2}{4\pi\,g}-\frac{K}{16\,\pi^2\,g^2}+\mathcal{O}(g^{-3})\Big)& 
\,.
\end{flalign}
Furthermore, with the same action it is possible to study perturbatively the (non-relativistic) dispersion relation for the field excitations over the classical string surface.  
For example, the corrections to the masses of the bosonic fields $x,x^*$ in \eqref{S_cusp}  (defined as the values of energy at vanishing momentum)  read~\cite{Giombi:2010bj}
%Considering  e.g. the bosonic fields $x,x^*$ in \eqref{S_cusp},  
%the corresponding corrections to their masses  reads~\cite{Giombi:2010bj}
\begin{equation}\label{mx}
m^2_{ x}(g)=\frac{m^2}{2}\,\Big(1-\frac{1}{8 \,g}+\mathcal{O}(g^{-2})\Big)~,
\end{equation}
where, as mentioned above, we restored the dimensionful parameter $m$.  Both \eqref{cuspperturbative} and \eqref{mx} are results obtained in a dimensional regularization scheme in which power divergent contributions are set to zero.
 In what follows, we will compute the lattice correlators of the fields $x,x^*$ so to study whether  our discretization changes the renormalization pattern above. 

\bigskip

While the bosonic part of~\eqref{S_cusp} can be easily discretized and simulated, Gra\ss mann-odd fields are either ignored (quenched approximation) or  formally integrated out, letting their determinant become part - via exponentiation in terms of pseudofermions, see \eqref{fermionsintegration} below - of the Boltzmann weight of each configuration in the statistical  ensemble. In the case of higher-order fermionic interactions -- as in \eqref{S_cusp}, where they are at most quartic -- this is possible via the introduction of auxiliary fields realizing a linearization. Following~\cite{Roiban}, one introduces $7$ auxiliary  fields, one scalar $\phi$ and a  $SO(6)$ vector field $\phi_M$, with the following Hubbard-Stratonovich transformation 
\begin{eqnarray}\label{HubbardStratonovich}
&& \!\!\!\!\!\!\!
\exp \Big\{-g\int dt ds  \Big[-\textstyle{\frac{1}{{z}^{2}}}\left( {\eta}^{i}{\eta}_{i}\right)^{2}  +\Big(\textstyle{\frac{i}{ {z}^{2}}} {z}_{N} {\eta}_{i}{\rho^{MN}}_{\phantom{i}j}^{i} {\eta}^{j}\Big)^{2}\Big]\}\\\nonumber
&& 
\sim ~\int D\phi D\phi^M\,\exp\Big\{-  g\int dt ds\,[\textstyle\frac{1}{2}{\phi}^2+\frac{\sqrt{2}}{z}\phi\,\eta^2 +\frac{1}{2}({\phi}_M)^2-i\,\frac{\sqrt{2}}{z^2}\phi^M \,\big(\textstyle{\frac{i}{ {z}^{2}}} {z}_{N} {\eta}_{i}{\rho^{MN}}_{\phantom{i}j}^{i} {\eta}^{j}\big)]\Big\}~.
\end{eqnarray}
%where in the second line the contribution of a Jakobian is understood to occur.  
Above, in the second line we have  written  the Lagrangian for $\phi^M$ so to emphasize that it has an imaginary part. Indeed, the bilinear form in round brackets   is hermitian
\begin{equation}
\!
\Big(i\,\eta_i {\rho^{MN}}^i{}_j \eta^j\Big)^\dagger=-i(\eta^j)^\dagger({\rho^{MN}}^i{}_j)^*(\eta_i)^\dagger
=-i \eta_j\,{\rho^{MN}}_i{}^j\,\eta^i=i\eta_j\,{\rho^{MN}}^j{}_i\,\eta^i%\equiv   \mathrm{i}  \eta_i {\rho^{MN}}^i{}_j \eta^j
\,,
\end{equation}
as follows from the properties of the $SO(6)$ generators \eqref{rhomatrices}.
Since the auxiliary vector field $\phi^M$ has real support, the  Yukawa-term for it sets \emph{a priori} a phase problem~\footnote{In other words, the second quartic interaction in \eqref{HubbardStratonovich} is the square of an hermitian object and comes  in the exponential as a ``repulsive'' potential. This has  the final effect of an imaginary part in the auxiliary Lagrangian, precisely as the $i\,b\,x$ in $e^{-\frac{b^2}{4a}}\sim \int dx \, e^{-a x^2+ i b x}$, with $b\in \mathbb{R}$. }, the  only  question being whether the latter is treatable via standard reweighting. Below we will see that this is not the case for small values of $g$, suggesting that a different setting (alternative linearization) should be provided to explore the full nonperturbative region.

After the transformation \eqref{HubbardStratonovich}, the Lagrangian reads  
\begin{eqnarray}\label{Scuspquadratic}
{\cal L} &=&  {| \partial_t {x} + {\frac{m}{2}}{x} |}^2 + \frac{1}{{ z}^4}{\big| \partial_s {x} -\frac{m}{2}{x} |}^2
+ (\partial_t {z}^M + \frac{m}{2}{z}^M )^2 + \frac{1}{{ z}^4} (\partial_s {z}^M -\frac{m}{2}{z}^M)^2
\cr
&+&\frac{1}{2}{\phi}^2 +\frac{1}{2}({\phi}_M)^2+\psi^T O_F \psi\,
\label{final_continuum_L}~,
\end{eqnarray}
 with  $\psi\equiv({\theta}^i, { \theta}_i, {\eta}^i, {\eta}_i)$ and 
%%%%%%%%%%%%%%%%%%%%%%%%%%%%%%%%%%%%%%%%%%%%%
\begin{eqnarray}\nonumber
O_F & =&\left(\begin{array}{cccc}
0 & i\partial_{t} & -\mathrm{i}\rho^{M}\left(\partial_{s}+\frac{m}{2}\right)\frac{{z}^{M}}{{z}^{3}} & 0\\
\mathrm{i}\partial_{t} & 0 & 0 & -\mathrm{i}\rho_{M}^{\dagger}\left(\partial_{s}+\frac{m}{2}\right)\frac{{z}^{M}}{{z}^{3}}\\
\mathrm{i}\frac{{z}^{M}}{{z}^{3}}\rho^{M}\left(\partial_{s}-\frac{m}{2}\right) & 0 & 2\frac{{z}^{M}}{{z}^{4}}\rho^{M}\left(\partial_{s}{x}-m\frac{{x}}{2}\right) & i\partial_{t}-A^T\\
0 & \mathrm{i}\frac{{z}^{M}}{{z}^{3}}\rho_{M}^{\dagger}\left(\partial_{s}-\frac{m}{2}\right) &\mathrm{i}\partial_{t}+A & -2\frac{{z}^{M}}{{z}^{4}}\rho_{M}^{\dagger}\left(\partial_{s}{x}^*-m\frac{{x}}{2}^*\right)
\end{array}\right)\\\nonumber
\\
%\end{flalign}
%where
%\begin{eqnarray}
\label{OF}
A  &=&\frac{1}{\sqrt{2}{z}^{2}}{\phi}_{M}\rho^{MN} {z}_{N}-\frac{1}{\sqrt{2}{z}}{\phi}\, +\mathrm{i}\,\frac{{z}_{N}}{{z}^{2}}\rho^{MN} \,\partial_{t}{z}^{M}~.
%,\\ 
%{B}_i{}^{ j} & =&\frac{1}{\sqrt{2}{z}^{2}}{\phi}_{M}{\rho^{MN}}_i{}^j{z}_{N}+\frac{1}{\sqrt{2}{z}}{\phi}\,\delta_i{}^j+ \mathrm{i}\frac{{z}_{N}}{{z}^{2}}{\rho^{MN}}_i{}^j\,\partial_{t}{z}^{M}~.
\end{eqnarray}
%%%%%%%%%%%%%%%%%%%%%%%%%%%%%%%%%%%%%%%%%%%%%
Notice that \eqref{Scuspquadratic} and the integration measure involve only the field $\psi$ and not its complex conjugate~\footnote{The vector $\psi$ in \eqref{Scuspquadratic} collects the 8 complex $\theta$ and $\eta$ in a formally ``redundant'' way which includes both the fields and their complex conjugates. Explicitating real and imaginary parts of $\theta,\eta$, it is easy to see that the fermionic contribution coming from this  $16\times 16$ complex operator $O_F$ is then the one of $16$ \emph{real} anti-commuting degrees of freedom.}, thus formally integrating out generates a Pfaffian ${\rm Pf}\,O_F$ rather than a determinant. In order to enter the Boltzmann weight and thus  be interpreted as a probability, ${\rm Pf}\,O_F$ 
should be  positive definite.
For this reason, we proceed as in~\cite{Roiban} 
\begin{equation}\label{fermionsintegration}
 \int \!\! D\Psi~ e^{-\textstyle\int dt ds \,\Psi^T O_F \Psi}={\rm Pf}\,O_F\equiv(\det O_F\,O^\dagger_F)^{\frac{1}{4}}= \int \!\!D\xi D\bar\xi\,e^{-\int dt ds\, \bar\xi(O_FO^\dagger_F)^{-\frac{1}{4}}\,\xi}~,
 \end{equation}
where the second equivalence obviously ignores potential phases or anomalies.

\section{Discretization}
\label{sec:discretization}

In order to investigate the lattice model corresponding to \eqref{Scuspquadratic}, we introduce a two-dimensional grid with lattice spacing $a$.  
We assign the values of the discretised (scalar) fields to each lattice site,  with periodic boundary conditions for all the fields except for antiperiodic temporal boundary conditions in the case of fermions. The discrete approximation of continuum derivatives are finite difference operators defined on the lattice. 
While this works well for the bosonic sector, 
a Wilson-like lattice operator must be introduced such that fermion doublers are suppressed.    
%~\footnote{\colb{At variance with~\cite{Roiban}, where a discretization of (doubler free) $O_F^\dagger O_F $ is proposed, we discretize $O_F$.}}.  
Due to the rather non-trivial structure of the Dirac-like operator in \eqref{OF} there are in principle many possible ways of introducing a Wilson-like operator. An optimal discretization should preserve all the symmetries of the continuum action and should lead to lattice perturbative calculations reproducing, in the $a\to0$ limit, the  continuum behavior \eqref{cuspperturbative}. Furthermore, in order not to prevent Montecarlo simulations the discretization should not induce complex phases in the fermionic determinant -- here, no complex phase should be added to the one already implicit in the Hubbard-Stratonovich procedure adopted. We will find  that it is not possible to satisfy all these requirements and therefore we choose to give up the global $U(1)$ symmetry. Let us discuss the procedure in details. For simplicity we start with the continuum model (reviewed in Appendix \ref{app:continuum}) and we denote with $u^M$ a particular $SO(6)$ direction (i.e. such that $u^M u_M=1$) defining the vacuum around which we expand the operator \eqref{OF} perturbatively (as an example, in \eqref{exp6}  $u^M=(0,0,0,0,0,1)$ has been chosen). 
The free, kinetic part of the fermionic operator \eqref{OF} in Fourier transform reads 
 \begin{equation}\label{fermkin}
K_F=\left(\begin{array}{cccc}
			0 			&- p_0 \mathbb{1}			& (p_1-i\frac m 2 ) \rho^M u_M	& 0 \\
			- p_0\mathbb{1}	&0					&0				& (p_1-i\frac m 2 )\rho_M^\dagger u^M\\
			-(p_1+i\,\frac m 2 ) \rho^M u_M	&0					&0				&- p_0 \mathbb{1}\\
			0			&-(p_1+i\,\frac m 2 )\rho_M^\dagger u^M	&- p_0\mathbb{1}		&0
          \end{array}\right)\,,\qquad
 \end{equation}
and to compute its determinant one can use the block matrix identity
 \begin{equation}
\det K_F=\det \left(\begin{array}{cc}
			K_1 			& K_2 \\
			K_3	&K_4				
			\end{array}\right)=\det(K_1) \det(K_4-K_3 K_1^{-1} K_2)=\det(K_4) \det(K_1-K_2 K_4^{-1} K_3)
\end{equation}
The simplicity of the matrix $K_3 K_1^{-1} K_2$ (or, equivalently $K_2 K_4^{-1} K_3$)
\begin{equation}
 K_3 K_1^{-1} K_2=\left(\begin{array}{cc}
			0 			& -\frac{i \left(m^2+4 p_1^2\right)}{4 p_0} \mathbb{1} \\
			-\frac{i \left(m^2+4 p_1^2\right)}{4 p_0}\mathbb{1}	&0				
			\end{array}\right)			
\end{equation}
immediately shows that 
\be\label{freefermdet}
 \det K_F=\Big(p_0^2+p_1^2+ \frac{m^2}{4}\Big)^8~.
\ee
From this result it is immediate to realize that for the fermionic degrees of freedom  the naive discretization~\cite{montvay} 
\be\label{naive}
p_\mu\to \mathring{p}_\mu \equiv \frac{1}{a} \sin(p_\mu a)
\ee
gives rise to fermion doublers~\footnote{The doubling phenomenon corresponds  to the denominator of the fermionic  propagator   vanishing on the lattice not only for $p^2$ equal to the physical mass,
but also in other $2^d-1$ (here three) points -- the ones which have at least one component equal to $\pi/a$  and all the others vanishing. Fermionic propagators are here proportional to the relevant entries of the inverse of the fermionic kinetic operator \eqref{fermkin}.}. Notice that the vanishing entries in \eqref{fermkin} are set to zero by the $U(1)$ symmetry, as they couple fermions with the same charge. A $U(1)$-preserving  discretization should not affect those entries of the fermionic matrix, and should act only on the non-vanishing entries. Furthermore $SO(6)$ symmetry fixes completely the structure of the matrix \eqref{fermkin} so that the only Wilson term preserving all the symmetries would be of the form $p_0\to p_0+a_i$ and $p_1\to p_1 + b_i$ for different $a_i$ and $b_i$ in the four entries where $p_0$ and $p_1$ appear in \eqref{fermkin}. Implementing such a shift and computing the determinant of the fermionic operator one immediately finds that this would not yield the perturbative result \eqref{cuspperturbative} for any value of $a_i$ and $b_i$. Therefore we choose to break $U(1)$ symmetry and  introduce  the following  Wilson-like lattice operator
\begin{equation}\label{KF1}
{\hat K_F}=\left(\begin{array}{cccc}
      W_+      &- \mathring{p_0} \mathbb{1}      & (\mathring{p_1}-i\frac{m}{2} )\rho^M u_M  & 0 \\
      - \mathring{p_0}\mathbb{1}  &-W_+^{\dagger}          &0        & (\mathring{p_1}-i\frac{ m}{2})\rho_M^\dagger u^M\\
      -( \mathring{p_1}+i\frac{m}{2} )\rho^M u_M  &0          &W_-        &- \mathring{p_0} \mathbb{1}\\
      0      &-( \mathring{p_1}+i\,\frac{m}{2})\rho_M^\dagger u^M  &- \mathring{p_0}\mathbb{1}    &-W_-^\dagger
          \end{array}\right)~.
\end{equation}
where
\be\label{Wilsonshift1}
W_\pm = \frac{r}{2}\,\big({\hat p}_0^2\pm i\,{\hat p}_1^2\big)\, \rho^M u_M \,,
\ee
with $|r|=1$, and~\cite{montvay}  
%where $M=\frac12 a m$, 
\be\label{phat}
\hat p_\mu\equiv \frac{2}{a} \sin\frac{p_\mu a}{2}\,.
\ee
The analogue of \eqref{freefermdet} reads now
\be\label{detfermwilson}
\det  {\hat K}_F=\Big(\mathring p_0^2+\mathring p_1^2+\frac{r^2}{4} \left(\hat p_0^4+\hat p_1^4\right)+\frac{M^2}{4}\Big)^8
\ee
and can be used together with its bosonic counterpart -- obtained via the  naive replacement $p_\mu\to\hat p_\mu$ in the numerator of the ratio \eqref{oneloopcontinuum} -- to define in this discretized setting  the one-loop   partition function
\be\label{oneloopdiscrete}
\Gamma^{(1)}_{\rm LAT}=-\ln Z^{(1)}_{\rm LAT}=\mathcal{I}(a)
\ee
where, explicitly, for an infinite lattice
\begin{eqnarray}\nonumber
\mathcal{I}(a)=\frac{V_2}{2\,a^2}\!\!
\int\limits_{-\pi}^{+\pi}\!\!\frac{d^2p}{(2\pi)^2}
\ln\!\Big[\frac{4^8  (\sin ^2 \frac{p_0}{2} +\sin ^2 \frac{p_1}{2})^5 (\sin ^2 \frac{p_0}{2} +\sin ^2 \frac{p_1}{2}+\frac{M^2}{8})^2  (\sin ^2 \frac{p_0}{2} +\sin ^2 \frac{p_1}{2}+\frac{M^2}{4})}{\big(4 \sin ^4 \frac{p_0}{2} +\sin ^2p_0+4 \sin ^4 \frac{p_1}{2} +\sin ^2p_1+\frac{M^2}{4}\big)^8}\Big]\\
\label{integral}
\end{eqnarray}
and the integral above has been obtained rescaling the momenta with the lattice spacing and  setting $r=1$. 
A consistent discretization will be the one  for which  \eqref{oneloopdiscrete}-\eqref{integral}
converge in the $a\to 0$ limit to the  value in the continuum \eqref{oneloopcontinuum}. 
The integral \eqref{integral} can be indeed quickly performed numerically, leading to
\be\label{3log2}
\Gamma^{(1)}=-\ln Z^{(1)}=\lim_{a\to0}\mathcal{I}(a)=-\frac{3\ln 2}{8 \pi}\,N^2 M^2~\,,
\ee
where we used that $V_2= L^2 = (N a)^2$. 
Namely, expanding the integrand in \eqref{integral} around  $a\sim 0$  (recall that $M=m\,a$)
%~\footnote{As standard in perturbation theory, one proceeds considering a lattice of infinite volume ($L\to\infty$), from which the integral in \eqref{integral}.   Therefore, taking the continuum limit $a\to 0$ while keeping constant the physics -- which at finite volume $V=L^2$ we translate with $ m\,L=\text{const}$ as in \eqref{constantphysics} -- means here  keeping constant the parameter $m$ in $M=m\,a$.}) 
the $\mathcal{O}(a^0)~\text{and}~\mathcal{O}(a^1) $ terms vanish.  Then, with this discretization the cancellation in $\mathcal{I}(a)$ of quadratic $\sim\frac{1}{a^2}$ and linear $\sim \frac{1}{a}$ divergences (which in the continuum are related  to the equal number of fermionic and bosonic degrees of freedom and to the mass-squared sum rule) is ensured. The $\mathcal{O}(a^2)$ term provides then the continuum expected finite part. 

Given the structure of the Wilson term in the vacuum it is quite natural to generalize the prescription to the interacting case.
The discretized fermionic operator reads
\begin{eqnarray}\nonumber
\!\!\!\!\!\!\!\!\!\!\!\!\!\!\!\!
{\hat O_F}&\!\!=\!\!&\left(\begin{array}{cccc}
    \!\!\!\!  W_+      & \!\!\!\! - \mathring{p_0} \mathbb{1}      &\!\!\!\! (\mathring{p_1}-i\frac{m}{2} )\rho^M \frac{z^M}{z^3}  & \!\!\!\! 0 \\
 \!\!\!\!     - \mathring{p_0}\mathbb{1}  & \!\! -W_+^{\dagger}          &\!\!\!\! 0        & \!\!\!\! \rho_M^\dagger(\mathring{p_1}-i\frac{ m}{2})\frac{z^M}{z^3} \\
    \!\!\!\!  -( \mathring{p_1}+i\frac{m}{2} )\rho^M \frac{z^M}{z^3}   &\!\!\!\! 0          &\!\!\!\! 2\frac{{z}^{M}}{{z}^{4}}\rho^{M}\left(\partial_{s}{x}-m\frac{{x}}{2}\right) +W_-     &\!\!\!\! - \mathring{p_0} \mathbb{1}-A^T\\
   \!\!\!\!   0      &\!\!\!\! -\rho_M^\dagger( \mathring{p_1}+i\,\frac{m}{2})\frac{z^M}{z^3}   &\!\!\!\! - \mathring{p_0}\mathbb{1}+A    &\!\!\!\! -2\frac{{z}^{M}}{{z}^{4}}\rho_{M}^{\dagger}\left(\partial_{s}{x}^*-m\frac{{x}}{2}^*\right) -W_-^\dagger
          \end{array}\right)\\
          \label{OFgen}
          \end{eqnarray}
          with
          \be\label{Wilsonshiftgen}
W_\pm = \frac{r}{2\,z^2}\,\big({\hat p}_0^2\pm i\,{\hat p}_1^2\big)\,\rho^M z_M\,,
\ee
where a factor $1/z^2$ is present, which appears   to be useful for stability in the simulations  (to clarify/justify this structure a two-loop calculation in lattice perturbation theory would be needed). 
As we said, together with the requirement that the resulting determinant (in combination with the bosonic contribution) should reproduce the number in \eqref{3log2}, one important point is that the discretization should not induce (additional) complex phases. Indeed, consider  the continuum fermionic operator obtained setting to zero in \eqref{OF} those auxiliary fields $\phi^M$ whose Yukawa-term is responsible for the phase problem. It is easy to check  that  it satisfies the properties (antisymmetry and a constraint which is reminiscent of the $\gamma_5$-hermiticity in lattice QCD~\cite{montvay})
\be\label{gamma5prop}
\big(O_F|_{ \phi^M=0}\big)^T=-\,O_F|_{ \phi^M=0}\,,\qquad\qquad\qquad \big(O_F|_{ \phi^M=0}\big)^\dagger=\Gamma_5\,\big(O_F|_{ \phi^M=0}\big)\, \Gamma_5
\ee
where $\Gamma_5$ is the following unitary, antihermitian matrix
\be
\Gamma_5=\left(\begin{array}{cccc}
			0 			&  \mathbb{1}			&0& 0 \\
			-\mathbb{1}	&0					&0				&0\\
			0&    0					&0				& \mathbb{1}\\
			0			&0	&-\mathbb{1}		&0
          \end{array}\right)\,,
\qquad   \Gamma_5^\dagger \Gamma_5=\mathbb{1}  \qquad \Gamma_5^\dagger=-\Gamma_5\,.
\ee
The properties \eqref{gamma5prop} are enough to ensure that $\det O_F|_{ \phi^M=0}$ is \emph{real} and \emph{non-negative}. %so that  \eqref{fermionsintegration} is free of ambiguities for $\hat O_F$. 
Requiring that the addition of Wilson terms in the discretization of the (full) fermionic operator should preserve \eqref{gamma5prop}  is one of the criteria leading to ${\hat O_F}$ in \eqref{OFgen}. This is indeed what happens, as can be checked both numerically  and analytically, confirming that the phase problem described in Section \ref{sec:phase} is only due to the Hubbard-Stratonovich transformation.

 To answer the question about how restricted the choice of  %the freedom in the definition of the 
 Wilson-like operator introduced in \eqref{KF1} is, %is unique and why we were lead to such a choice. 
 one can show that  starting from a generic $16 \times 16$ matrix shift $V$ such that $\hat K_F=K_F+V$ it is possible to impose a set of constraints singling out the structure \eqref{KF1}. Here we summmarize  these requirements:
 \begin{itemize}
 \item $SO(6)$ invariance;
  \item Antisymmetry $\hat K_F^T=-\hat K_F$;
  \item $\Gamma_5$-hermiticity $\hat K_F ^\dagger=\Gamma_5\,\hat K_F\, \Gamma_5$;
  \item Determinant of $\hat K_F$ equal to \eqref{detfermwilson};
  \item Block structure for the matrix $K_3 K_1^{-1} K_2=\left(\begin{array}{cc}
			M_1		& M_2 \\
			M_3	 & M_4				
			\end{array}\right)$ with $[M_1,M_2]=[M_1,M_3]=0$.
 \end{itemize}
The $SO(6)$ invariance constrains the matrix $V$ to have a $4\times4$-block structure constructed out of the available $SO(6)$-invariant structures: $\mathbb{1}, \rho^M u_M$ and $\rho_M^\dagger u^M$.  The second and third requirements, as already mentioned, prevent the appearance of an unwanted phase in the fermionic determinant.  The fourth condition allows to reproduce the next-to-leading order perturbative result and the last constraint has been added to fix completely the form of the shift matrix $V$. In principle this last condition could be relaxed, but since we were able to find a matrix $V$ satisfying all these constraints, there is no need to do so.  
 
In Appendix~\ref{app:altern_discretization} we present simulations obtained with another fermionic discretization -- see \eqref{OFgenbreak}-\eqref{Wilsonshiftgenbreak} --  consistent only with lattice perturbation theory performed around vacua coinciding with one of six cartesian coordinates $u^M\,,~~M=1,\cdots,6$ (and no general linear combination of them) it breaks explicitly the  $SO(6)$ invariance of the model (again, the $U(1)$ symmetry is broken down as in the previous case). It is interesting to mention that, at least in the range  of the couplings explored, measurements of the two observables of interest here -- the $x$-mass and the derivative of the cusp -- and for the phase appear not to be sensitive to the different discretization. 

\section{Simulations, continuum limit and the phase}
\label{sec:simulation}
 As discussed above, in the continuum model there are two ``bare'' parameters,  the  dimensionless coupling $g=\frac{\sqrt{\lambda}}{4\pi}$  and the mass scale $m$.  
%~\footnote{We recall that $m$ just represent the $p_+$ scale lost after the original gauge-fixing.}.  
In taking the continuum limit, the dimensionless physical quantities that it is natural to keep constant when $a\to0$ are the physical masses of the field excitations  rescaled with $L$, the spatial lattice extent. This is our line of constant physics in the bare parameter space.  For the example in \eqref{mx}, this means 
\begin{equation}\label{constantphysics}
L^2 \,m^2_{x}=\text{const}\,,\qquad~~\text{leading to}\qquad~~L^2 \,m^2\equiv (N M)^2 =\text{const}\,,
\end{equation}
where we defined the dimensionless $M=m\, a$ with the lattice spacing $a$.  

The second equation in \eqref{constantphysics} relies first on the hypothesis that $g$ is not (infinitely) renormalized~\footnote{ This supposition is somewhat supported, a posteriori, by our analysis of the (derivative of the) scaling function, which can be used as a definition of the renormalized coupling. As discussed in Section~\ref{sec:action},  occurring divergences in $S_{\rm LAT}$ can be consistently subtracted  showing an agreement with the continuum expectation, at least for the region of lattice spacings and couplings that we explore.}.
Second, one should investigate whether the  relation \eqref{mx},  and the analogue ones for the other fields  of the model, are still true in the discretized model - \emph{i.e.}  the physical masses undergo only a finite renormalization. In this case, at each fixed $g$ fixing $ L^2~m^2$ constant would be enough to keep the rescaled physical masses constant, namely   no tuning of the ``bare'' parameter $m$ would be necessary.  %The second equation in \eqref{constantphysics} results from the following consideration. If the relation \eqref{mx}, and the  ones for the other fields  of the model, were still true in the discretized model - \emph{i.e.} the physical masses undergo only a \emph{finite} renormalization - and under the assumption that $g$ is not renormalized  (which is what happens in the continuum), at each fixed $g$ fixing $m^2~ L^2$ constant is enough to keep $L^2 \,m^2_{\rm x}$ constant, namely   no tuning of the ``bare'' parameter $m$ is necessary.   \colb{To establish that this might be a meaningful way to take the continuum limit,  one would have to study fermionic correlator}
%/Users/valentinaforini/Desktop/Photo Booth Library/Pictures/Movie on 4-13-16 at 4.02 PM.mov
%
In the present study, we start by considering the example of bosonic $x,x^*$ correlators, where indeed we find no ($1/a$) divergence for the ratio $m_x^2/m^2$ -- see section \ref{sec:correlator} below -- 
%the left panel in Figure \ref{fig:corr&action}. 
and in the  large $g$  region that we investigate  
%in general is under control (i.e. where standard reweighting works, see below) 
the ratio considered approaches the expected continuum value $1/2$.   
Having this as hint, and because with the proposed discretization we have recovered in perturbation theory the one-loop cusp anomaly \eqref{3log2}, we assume  that in the discretized model no further scale but the lattice spacing $a$  is present.  Any  observable $F_{\rm LAT}$ is therefore a function 
of the  input (dimensionless)  ``bare''  parameters $g, N$ and $M$
\begin{equation}
F_{\rm LAT}=F_{\rm LAT}(g,N,M)=F(g)+\mathcal{O}\Big(\frac{1}{N}\Big)+\mathcal{O}\,\left(e^{- MN}\right)
\end{equation}
where
\begin{equation}
g=\frac{\sqrt{\lambda}}{4\pi}
\,,\qquad  N=\frac{L}{a}\,,\qquad M=a \,m\,.
\end{equation}
At fixed coupling $g$ and fixed $m\, L\equiv M\,N$ (large enough so to keep finite volume effects $\sim e^{-m\,L}$  small),    $F_{\rm LAT}$ is evaluated for different values of $N$ and it differs from its continuum equivalent by lattice artifacts $\mathcal{O}\big(\frac{1}{N}\big)$. The continuum limit $F(g)$  is obtained via an extrapolation to infinite~$N$. While most runs are done at $m\,L=4$, for one value of the coupling ($g=30$) we perform simulations at a larger value ($m\,L=6$, orange point in the continuum plots) to explicitly check finite volume effects. For the physical observables under investigation, Figs. \ref{fig:correlator} (right panel) and \ref{fig:action_fin_g} , we find these effects to be very small and within the present statistical errors. They appear to play a role only in the case of the coefficient of the divergences which must be subtracted non-perturbatively in order to define the cusp action, see Section \ref{sec:action}, as in Fig. \ref{fig:constant} (right panel).  
  \begin{figure}[t]
   \centering
   \includegraphics[scale=.7]{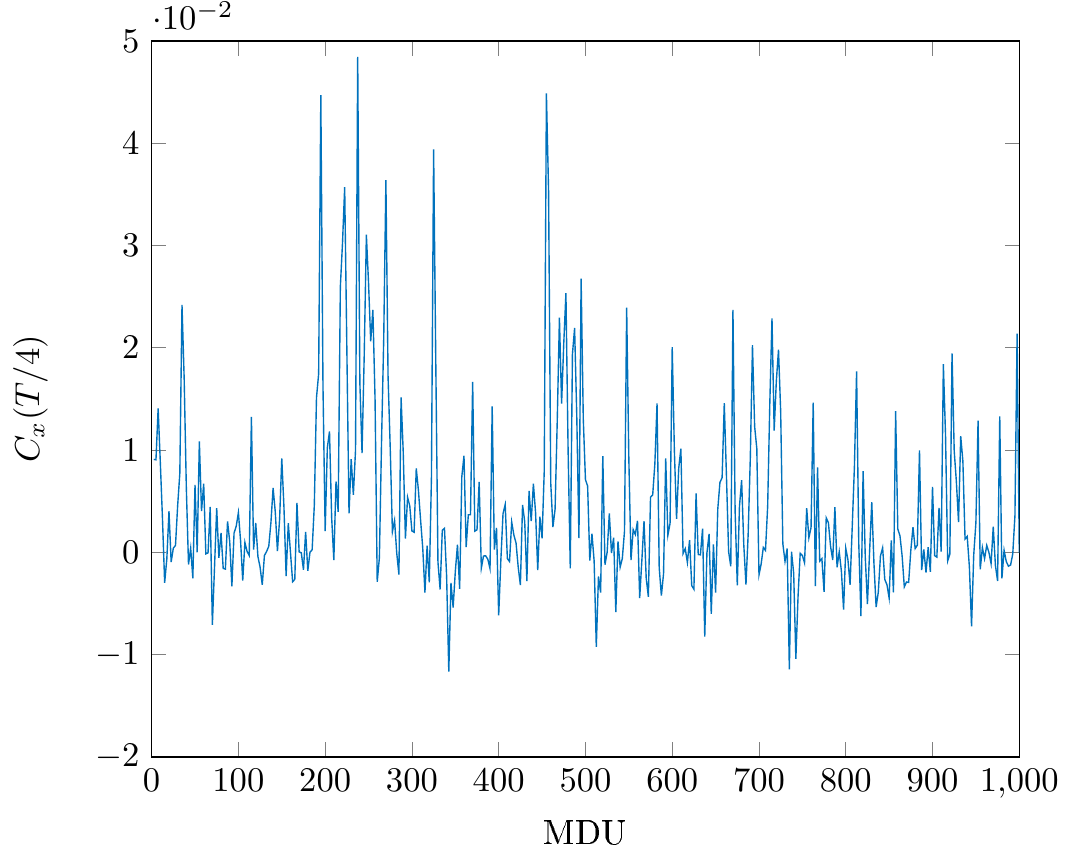}
    \includegraphics[scale=.7]{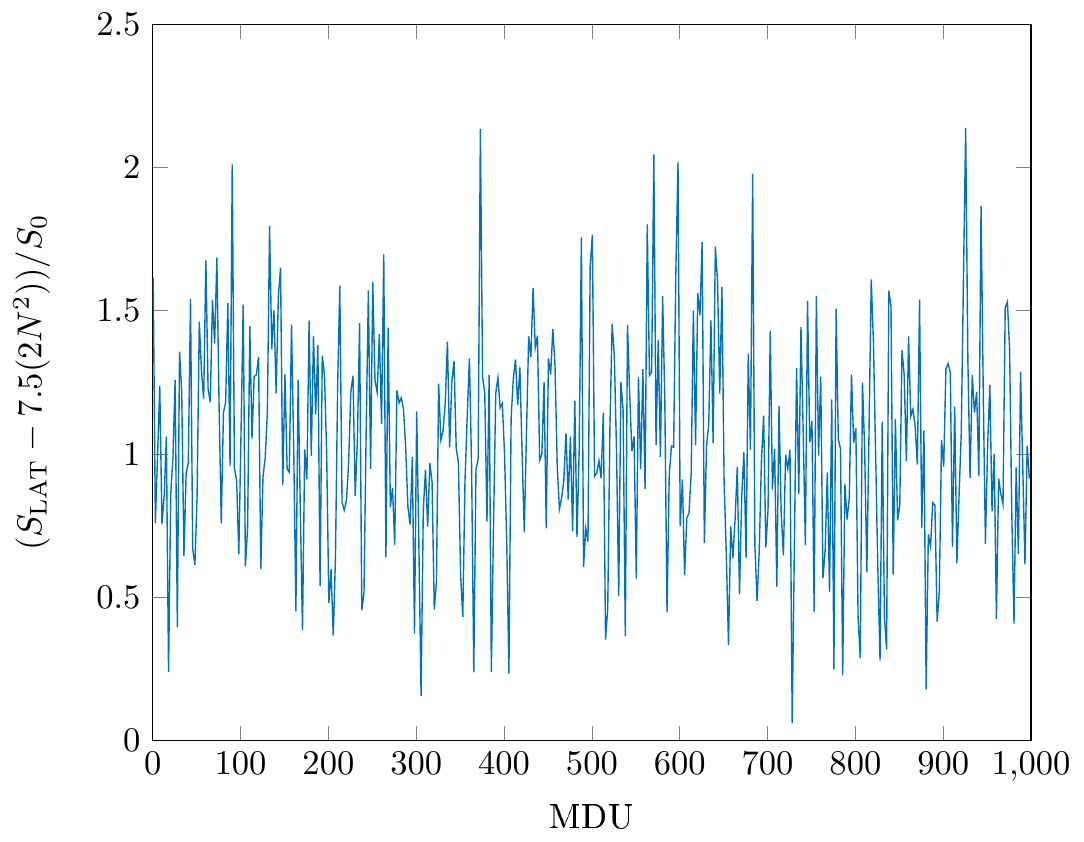}
 \caption{Monte Carlo histories for the  correlator $\langle x\,x^*\rangle$ at time separation $T/4$ and for $\langle S_{\rm cusp}\rangle$, at $g=10$ and $L/a=16$, in terms of Molecular Dynamic Units (MDU). 
The HMC produces a series of bosonic field configurations, on each of them the observable is evaluated and plotted here for the same series  at the given parameters. The fact that  successive configurations produced by the RHMC are statically correlated might lead to strong so-called auto-correlations in the data, which would appear in these plots as fluctuations with long periods. As one can see, the histories presented here do not suffer from such long fluctuations, and sample well the observables under investigation.}
\label{fig:MChistories}
\end{figure} 

\begin{table}[H]
\centering
\begingroup
\renewcommand*{\arraystretch}{0.9}
\begin{tabular}{ccccccc}
\toprule
$g$ & $T/a\times L/a$ & $Lm$ & $am$ & $\tau_{\rm int}^{S}$ & $\tau_{\rm int}^{m_x}$ & statistics [MDU] \\
\midrule
  5 & $16 \times   8$ &  4  & 0.50000 & 0.8 & 2.2 & 900 \\
    & $20 \times  10$ &  4  & 0.40000 & 0.9 & 2.6 & 900 \\
    & $24 \times  12$ &  4  & 0.33333 & 0.7 & 4.6 & 900,1000 \\
    & $32 \times  16$ &  4  & 0.25000 & 0.7 & 4.4 & 850,1000 \\
    & $48 \times  24$ &  4  & 0.16667 & 1.1 & 3.0 & 92,265 \\
\midrule
 10 & $16 \times   8$ &  4  & 0.50000 & 0.9 & 2.1 & 1000 \\
    & $20 \times  10$ &  4  & 0.40000 & 0.9 & 2.1 & 1000 \\
    & $24 \times  12$ &  4  & 0.33333 & 1.0 & 2.5 & 1000,1000 \\
    & $32 \times  16$ &  4  & 0.25000 & 1.0 & 2.7 & 900,1000 \\
    & $48 \times  24$ &  4  & 0.16667 & 1.1 & 3.9 & 594,564 \\
\midrule
 20 & $16 \times   8$ &  4  & 0.50000 & 5.4 & 1.9 & 1000 \\
    & $20 \times  10$ &  4  & 0.40000 & 9.9 & 1.8 & 1000 \\
    & $24 \times  12$ &  4  & 0.33333 & 4.4 & 2.0 & 850 \\
    & $32 \times  16$ &  4  & 0.25000 & 7.4 & 2.3 & 850,1000 \\
    & $48 \times  24$ &  4  & 0.16667 & 8.4 & 3.6 & 264,580 \\
\midrule
 30 & $20 \times  10$ &  6  & 0.60000 & 1.3 & 2.9 & 950 \\
    & $24 \times  12$ &  6  & 0.50000 & 1.3 & 2.4 & 950 \\
    & $32 \times  16$ &  6  & 0.37500 & 1.7 & 2.3 & 975 \\
    & $48 \times  24$ &  6  & 0.25000 & 1.5 & 2.3 & 533,652 \\
    & $16 \times   8$ &  4  & 0.50000 & 1.4 & 1.9 & 1000 \\
    & $20 \times  10$ &  4  & 0.40000 & 1.2 & 2.7 & 950 \\
    & $24 \times  12$ &  4  & 0.33333 & 1.2 & 2.1 & 900 \\
    & $32 \times  16$ &  4  & 0.25000 & 1.3 & 1.8 & 900,1000 \\
    & $48 \times  24$ &  4  & 0.16667 & 1.3 & 4.3 & 150 \\
\midrule
 50 & $16 \times   8$ &  4  & 0.50000 & 1.1 & 1.8 & 1000 \\
    & $20 \times  10$ &  4  & 0.40000 & 1.2 & 1.8 & 1000 \\
    & $24 \times  12$ &  4  & 0.33333 & 0.8 & 2.0 & 1000 \\
    & $32 \times  16$ &  4  & 0.25000 & 1.3 & 2.0 & 900,1000 \\
    & $48 \times  24$ &  4  & 0.16667 & 1.2 & 2.3 & 412 \\
\midrule
100 & $16 \times   8$ &  4  & 0.50000 & 1.4 & 2.7 & 1000 \\
    & $20 \times  10$ &  4  & 0.40000 & 1.4 & 4.2 & 1000 \\
    & $24 \times  12$ &  4  & 0.33333 & 1.3 & 1.8 & 1000 \\
    & $32 \times  16$ &  4  & 0.25000 & 1.3 & 2.0 & 950,1000 \\
    & $48 \times  24$ &  4  & 0.16667 & 1.4 & 2.4 & 541 \\
\bottomrule
\end{tabular} 
\endgroup
\caption{
Parameters of the simulations: the coupling $g$, the temporal ($T$) and spatial ($L$) extent
of the lattice in units of the lattice spacing $a$, the line of constant physics fixed by
$Lm$ and the mass parameter $M=am$. The size of the statistics after thermalization is given
in the last column in terms of Molecular Dynamic Units (MDU), which equals an HMC trajectory
of length one. In the case of multiple replica the statistics for each replica is
given separately. The auto-correlation times $\tau$ of our main observables $m_x$ and $S$ are also given
in the same units.
}
\label{t:runs}
\end{table}
In Table \ref{t:runs}  we list the parameters of the simulations presented in this paper. 
The Monte Carlo evolution of each $F_{\rm LAT}(g,N,M)$   is generated by the standard Rational Hybrid Monte Carlo (RHMC) algorithm~\cite{RHMC1,RHMC2}.  The rational approximation  for the inverse fractional power in the last equation of \eqref{fermionsintegration} %--  as in \cite{Roiban}. The rational of approximation 
is of degree $15$, and we checked for a subset of the configurations that its  accuracy is always better than $10^{-3}$ for $\bar {\xi}(O_F O_F^+)^{-1/4}\xi$. %for the pseudo-fermionic action.
In Fig.\ref{fig:MChistories} we show examples of Monte Carlo histories for our two main observables - the correlator $\langle x^* x\rangle$  and the action $\langle S_{\rm cusp}\rangle$. 
We determined  auto-correlation times of the observables and included their effect in the error analysis~\cite{Wolff:2003sm}. 
Multiple points at the same value of $g$ and $N$ in Fig. \ref{fig:correlator} (left panel), Fig. \ref{fig:action_div} and Fig. \ref{fig:action_fin_N2}
-- and similarly in Fig. \ref{fig:correlator_break} (left panel), Fig. \ref{fig:action_div_break} and Fig. \ref{fig:action_fin_N2_break} -- indicate multiple replica.

\subsection{The $\langle x\,x^*\rangle$ correlator}
\label{sec:correlator}

To motivate the line of constant physics \eqref{constantphysics}, we investigate in this section  the physical mass of the bosonic fluctuation field $x$ around the string vacuum \eqref{null_cusp_back} as determined from the $\langle x\,x^*\rangle$ correlator.  The masses of the bosonic fields $x,x^*$ in \eqref{S_cusp}  (defined as the values of energy at vanishing momentum) can be read off, at leading order, from the expansion of the quadratic fluctuation Lagrangian \eqref{quadraticlagr}. The leading quantum correction to their dispersion relation have been computed in~\cite{Giombi:2010bj}, leading to~\eqref{mx}~\footnote{The prediction for the whole spectrum of excitations  was obtained via asymptotic Bethe ansatz in \cite{Basso:2010in} and later confirmed by semiclassical string theory around the folded closed string in $AdS_5$ in the large spin limit \cite{Giombi:2010bj}. 
The world surface spanned by the latter is equivalent~\cite{Kruczenski:2007cy}, via an analytic continuation and a global $SO(2,4)$ transformation, to that of the null cusp solution \eqref{null_cusp_back}. Notice that the mass spectrum in light-cone gauge coincides with the one in conformal gauge up to a factor of 4~\cite{Giombi}.}. 
%  We are only interested in the physical mass $m_x^{\textrm{phys}}(g)$, \emph{i.e.} the value of the excitation energy at vanishing spatial momentum, at large $g$~\footnote{Notice that the mass spectrum in light-cone gauge coincides with the one in conformal gauge up to a factor of 4 \cite{Giombi} and we reinstated the mass parameter $m$.}
%
\begin{figure}[t]
    \centering
           \includegraphics[scale=0.65]{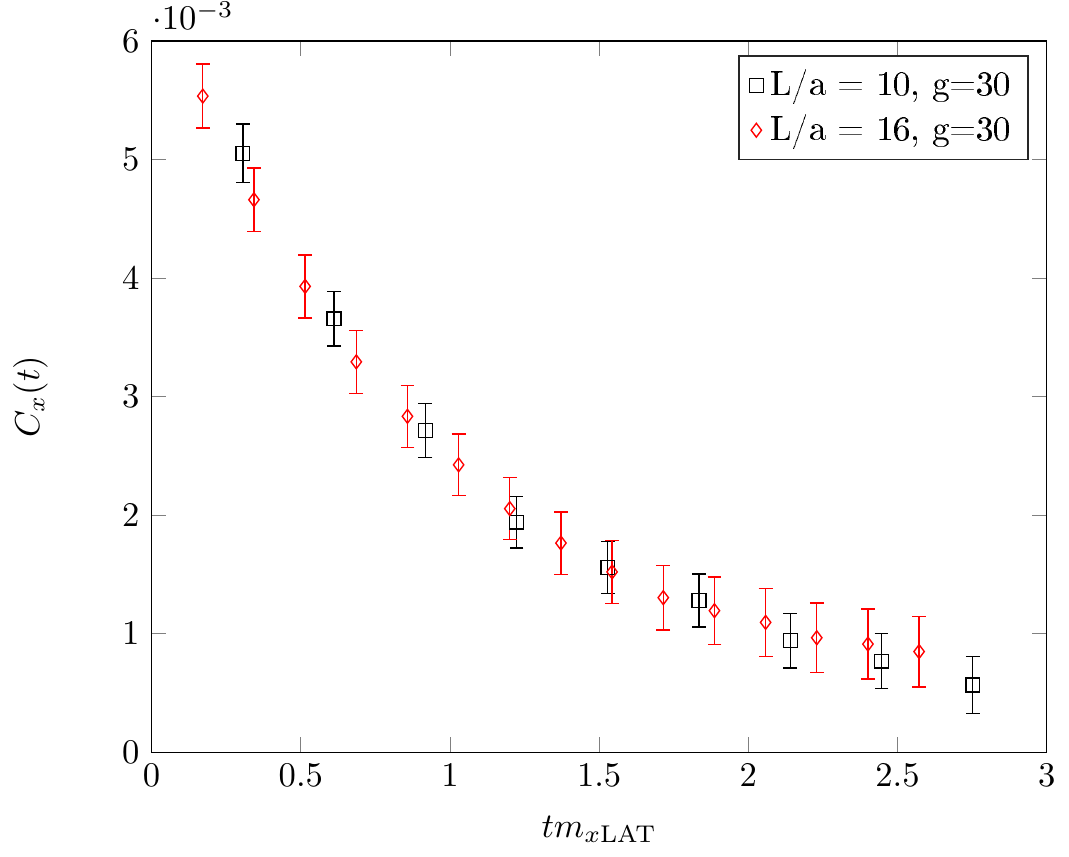}        
         \hspace{0.5cm}
         %  \caption{ } 
         \hspace{0.5cm}
          \includegraphics[scale=0.65]{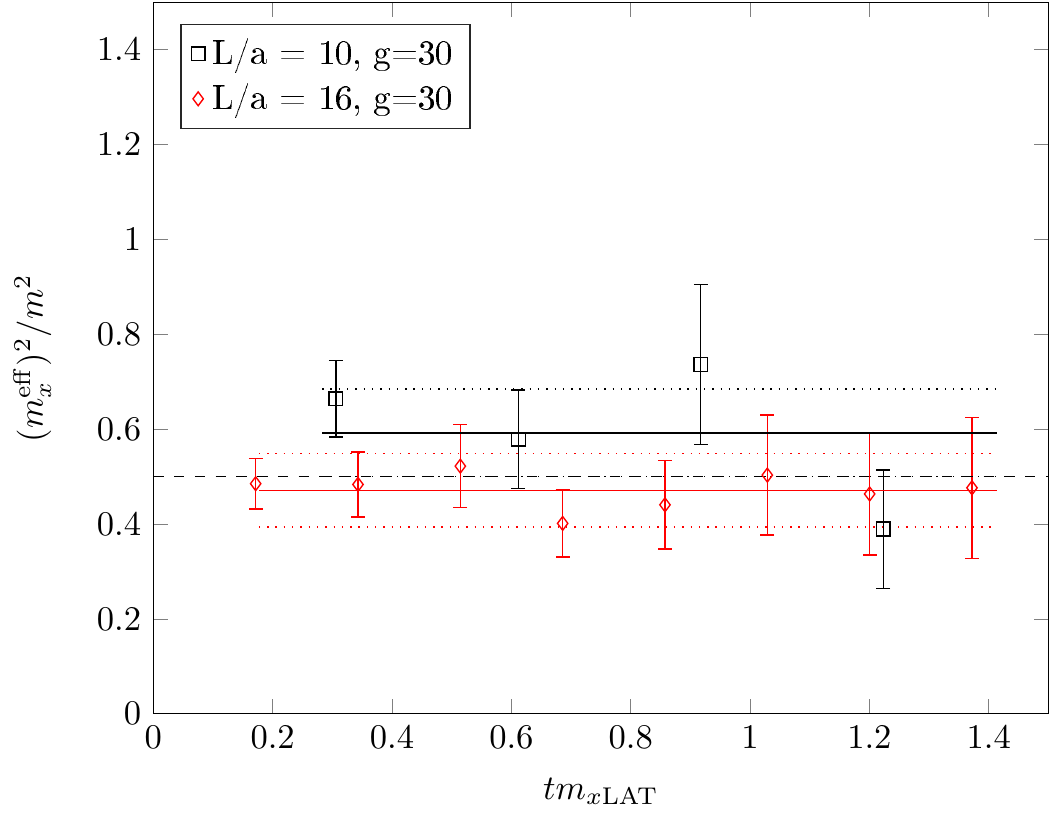}
           \hspace{0.5cm}
            \caption{Correlator $C_x(t)=\sum_{s_1,s_2} \langle x(t,s_1) x^*(0,s_2)\rangle$ of bosonic fields $x,x^*$ ({\bf left panel}) and corresponding effective mass $m^{\rm eff}_x=\frac{1}{a}\ln\frac{C_x(t)}{C_x(t+a)}$ normalized by $m^2$ ({\bf right panel}), plotted as functions of the time $t$ in units of ${m_x}_{\rm LAT}$ for different $g$ and lattice sizes. The flatness of the effective mass indicates that the ground state saturates the correlation function, and allows for a reliable extraction of the mass of the $x$-excitation.  %Exponentially decaying corrections to the observed constant value for the ratio  (expected at small values of $t\,m_x$ and  due to excited states) appear to be absent.
            Data points are masked by large errorbars for time scales greater than unity because the signal of the correlator degrades exponentially compared with the   statistical noise. } 
             \label{fig:correlatormass}
\end{figure}
\begin{figure}[h]
    \centering
        %  \caption{ Effective mass plot $m^{\rm eff}_x=\frac{1}{a}\ln\frac{C_x(t)}{C_x(t+a)}$, as calculated from the  correlator $C_x(t)=\sum_{s_1,s_2} \langle x(t,s_1) x^*(0,s_2)\rangle$ of bosonic fields $x,x^*$ in  presence  of Wilson terms.} 
                \includegraphics[scale=0.65]{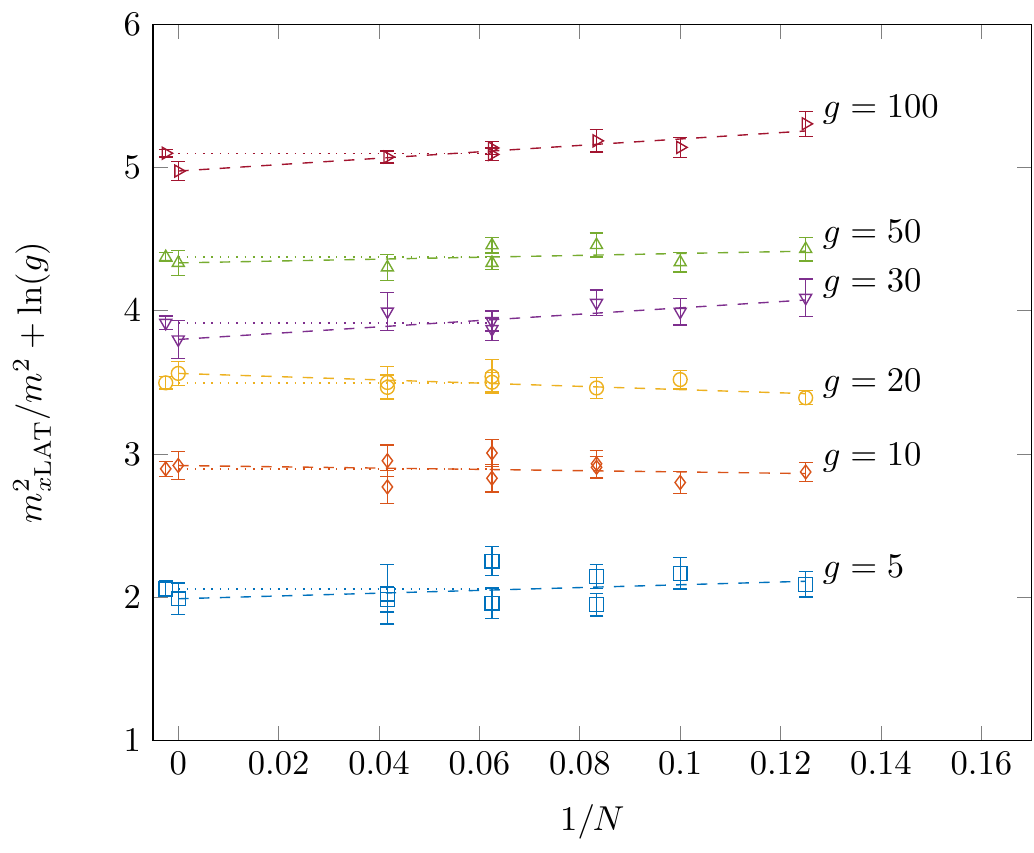}
\hspace{0.5cm}
              \hspace{0.5cm}
    \includegraphics[scale=0.65]{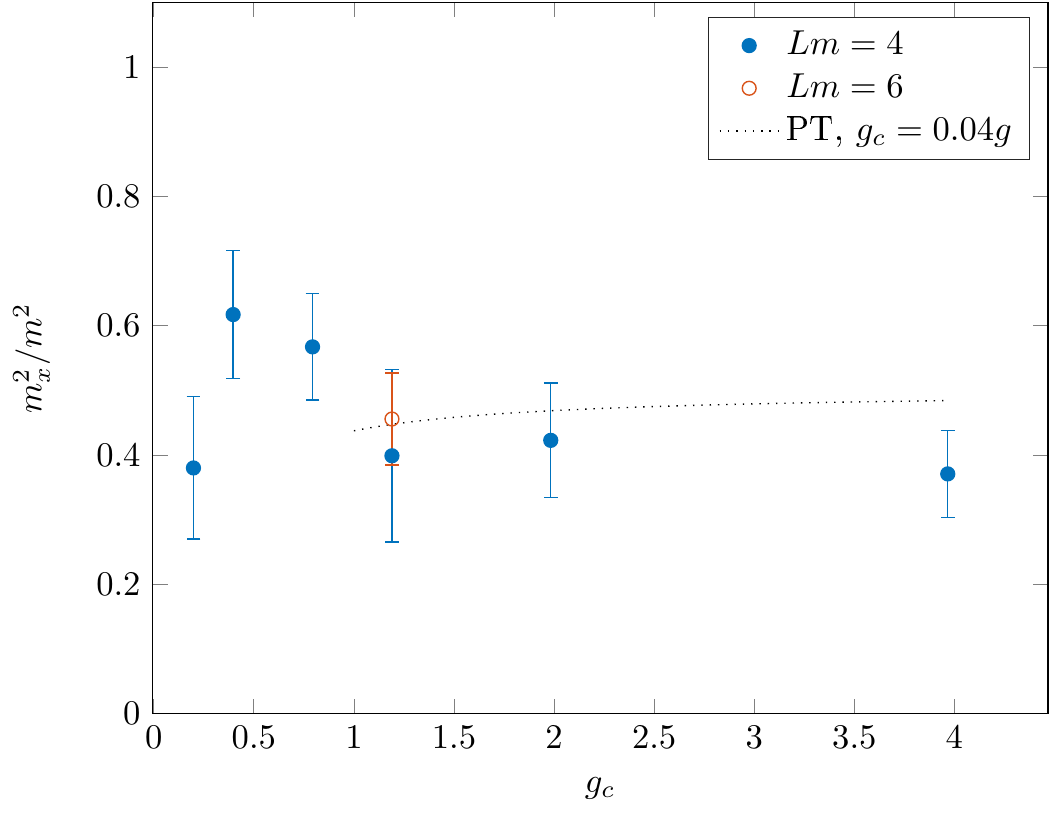}
      %  \caption{ Effective mass plot $m^{\rm eff}_x=\frac{1}{a}\ln\frac{C_x(t)}{C_x(t+a)}$, as calculated from the  correlator $C_x(t)=\sum_{s_1,s_2} \langle x(t,s_1) x^*(0,s_2)\rangle$ of bosonic fields $x,x^*$ in  presence  of Wilson terms.} 
                    \caption{{\bf Left panel}: 
                   Plot of $m^2_{\rm x LAT}(N,g)/m^2 = m_x(g) + \mathcal{O}(1/N)$, as from plateaux average of results which for $g=30$ are shown in Fig. \ref{fig:correlatormass} (right panel). To ensure better visibility of the fits at different $g$ values, $\ln g$ has been added.  Dashed lines represent a linear fit to all the data points for one value of $g$, while for dotted lines the fit is to a constant and only includes the two smallest lattice spacings. Multiple points at the same value of $g$ and $N$ indicate multiple replica. {\bf Right panel}: Continuum extrapolation corresponding to the linear fits in the left panel. The simulations represented by the orange point ($m\,L=6$) are used for a check of the finite volume effects, that appear to be within statistical errors. The extrapolation is plotted as a function of the continuum coupling $g_c=0.04\, g$ to facilitate the comparison with the prediction coming from the perturbative expectation (PT) \eqref{mx}, and uses the matching procedure performed for the observable action. The latter is described in Section \ref{sec:action} and commented further in Section \ref{sec:conclusions}. 
                                     } 
            \label{fig:correlator}
\end{figure}
One can estimate the dependence of the physical mass on the coupling constant by measuring the connected two-point correlation function of the discretised $x$-field on the lattice (see for example~\cite{montvay}).  
%As the magnitude of the first correction is too small to be detected in the range $g>10$ attainable by the simulations, we only attempt to match the classical constant behaviour. 
In configuration space one defines the two-point function
\begin{eqnarray}
\label{corr_funct_def}
G_x(t_1,\,s_1;\,t_2,\,s_2)&=&\langle x(t_1,\,s_1) x^* (t_2,\,s_2) \rangle\,
\end{eqnarray}
and Fourier-transforms over spatial directions to define the lattice timeslice correlator 
\begin{gather}\label{timeslicecorr}
C_x(t;\,k)=\sum_{s_1,\,s_2} e^{-i k (s_1-s_2)} G_x(t,\,s_1;\,0,\,s_2)\:.
\end{gather}
%Translational invariance of the field and its boundary conditions allowed to set one temporal position to zero. This form is particularly useful to study the energy spectrum of the theory. 
The latter 
%In fact, the partially Fourier transformed correlator 
admits a spectral decomposition over propagating states of different energies, given spatial momentum $k$ and amplitude $c_n$
\begin{gather}
C_x(t;\,k) = \sum_{n} |c_n|^2 e^{- t E_x(k;\,n)}
\end{gather}
which is dominated by the state of lowest energy for sufficiently large temporal distance $t$. 
This effectively single asymptotic exponential decay corresponds to a one-particle state with energy equal - for vanishing spatial momentum - to the physical mass of the $x$-field 
\begin{gather}\label{correlator_groundstate}
C_x(t;\,0)~ \stackrel{t\gg1}{\sim} ~e^{- t\, {m_x}_{\rm LAT}}, \qquad \qquad {m_x}_{\rm LAT}=E_x(k=0)~.
\end{gather}
On the lattice, the physical mass ${m_x}_{\rm LAT}$ is usefully obtained as a limit of an \emph{effective mass} $m^\textrm{eff}_{x}$, defined at a given timeslice extension $T$  and fixed timeslice pair $(t,t+a)$ by the discretized logarithmic derivative of the  timeslice correlation function \eqref{timeslicecorr} at zero momentum
\begin{gather}
\label{measured_m_eff}
{m_x}_{\rm LAT}=\lim_{T,\,t\to\infty}m^\textrm{eff}_{x}\equiv \lim_{T,\, t\to\infty, }\frac{1}{a} \log\frac{C_x(t;\,0)}{C_x(t+a;\,0)}.
\end{gather}
Figure \ref{fig:correlatormass} shows the effective mass measured from \eqref{measured_m_eff}  as a function of the time $t$ in units of ${m_x}_{\rm LAT}$ for different $g$ and lattice sizes. To reduce uncertainty about the saturation of the ground state in the correlation function - in \eqref{correlator_groundstate}, corrections to the limit are proportional to $e^{-\Delta E\, t}$, where $\Delta E$ is the energy splitting with the nearest excited state -- in our simulations the lattice temporal extent $T$ is always twice the spatial extent $L$. 
 The flatness of the effective mass in Fig. \ref{fig:correlatormass} (right) indicates that the ground state saturates the correlation function, and allows for a reliable extraction of the mass of the $x$-excitation.  Data points are masked by large errorbars for time scales greater than unity because the signal in \eqref{measured_m_eff} degrades exponentially compared with the   statistical noise. Our simulations provide an estimate for the $x$ mass, $m_{x}^2 / m^2 =\frac{1}{2}$   that appears to be consistent with the classical, large $g$ prediction \eqref{mx}. We do not see a clear signal yet for the expected bending down at smaller $g$. For decreasing couplings simulations become compute-intensive and to obtain smaller errors longer/parallel runs would be necessary. 

The most important corollary of  the analysis for the $\langle x x^*\rangle$ correlator is the following. As it happens in the continuum, also in the discretized setting there appears to be no infinite renormalization occurring for \eqref{mx}, and thus no need of tuning the bare parameter $m$ to adjust for it. 
This corroborates the choice of \eqref{constantphysics} as the line of constant physics along which a continuum limit can be taken.  

\subsection{The cusp action}
\label{sec:action}

%In measuring  the action \eqref{vevaction} on the lattice, we are supposed to recover the following general behavior 
%\begin{equation} \label{fit}
%\frac{\langle S_{\rm LAT}\rangle}{N^2} = \frac{c}{2}+\frac{1}{8}\,M^2\,g \,f'(g)\,, 
%\end{equation}
% where we have reinserted the parameter $m$, used  that $V_2=a^2\,N^2$ and added a constant  contribution  in $g$ which takes into account possible coupling-dependent Jacobians relating the (derivative of the) partition function on the lattice to the one in the continuum. 
 %~\footnote{In \cite{Roiban} this contribution has been subtracted via a suitable normalization.}. 

%
\begin{figure}[t]
    \centering
     \includegraphics[scale=0.68]{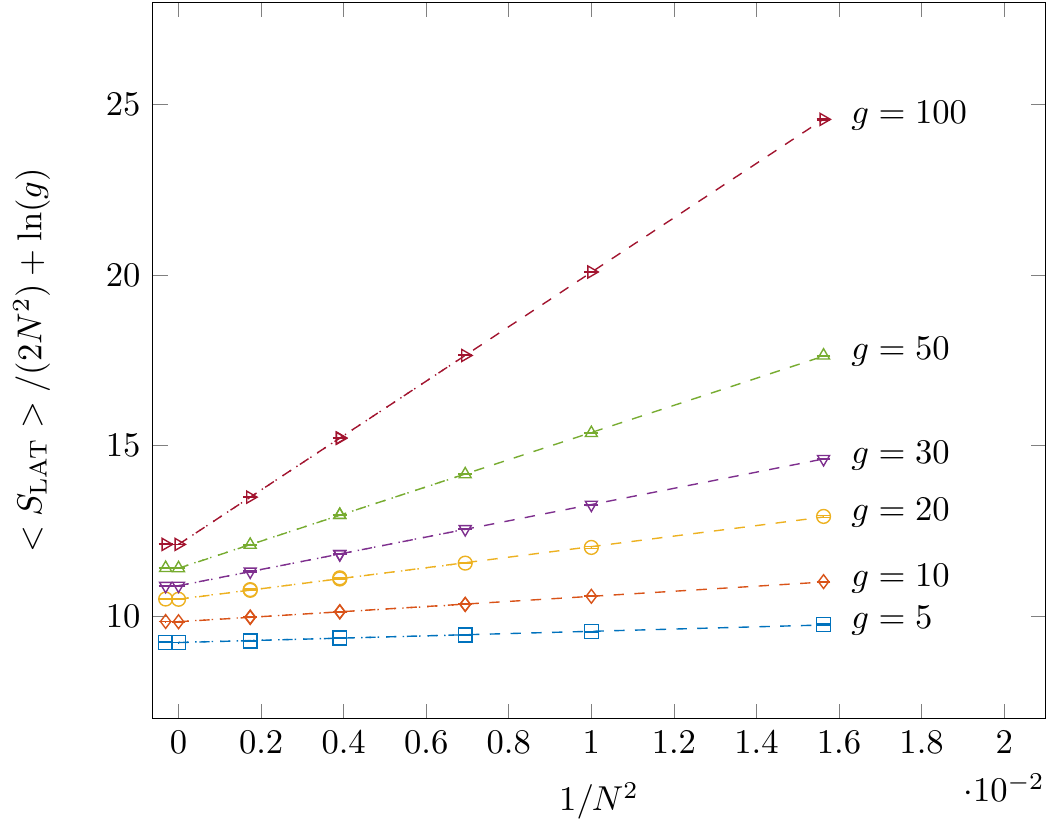}
      \hspace{0.5cm}
  \includegraphics[scale=0.71]{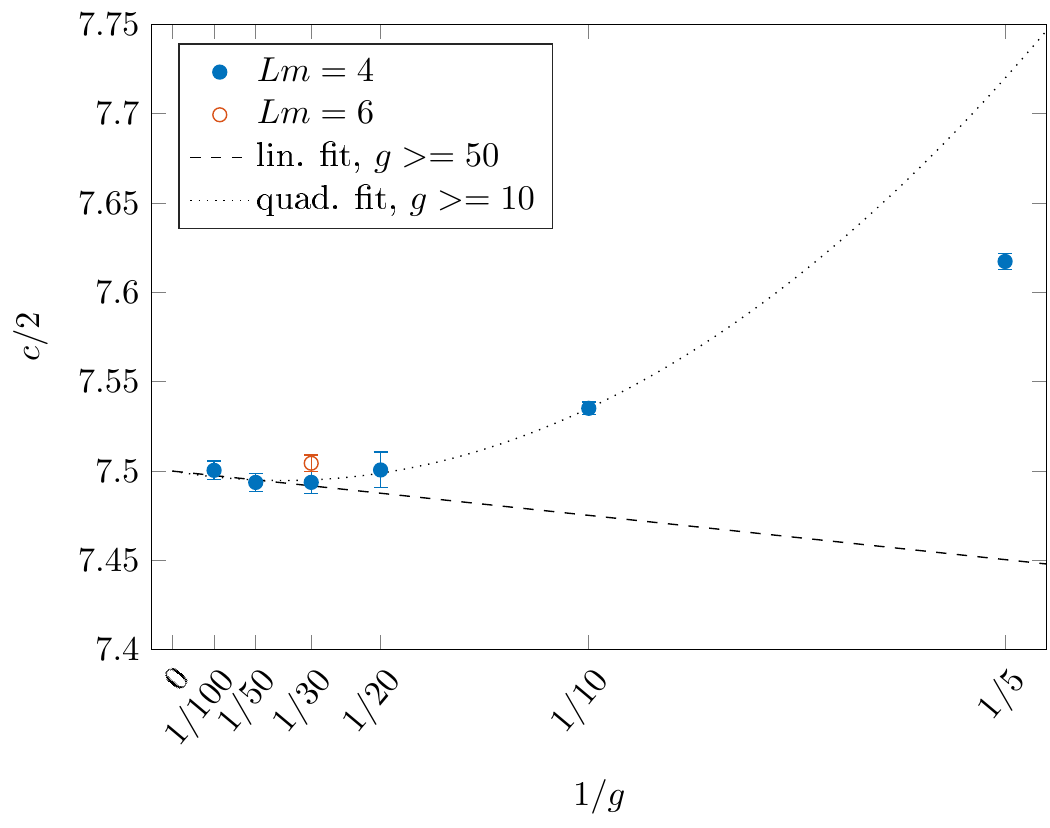}
        \caption{   {\bf Left panel}: Plots  of $\frac{\langle S_{\rm LAT}\rangle}{2 N^2}$, where fits  (dashed lines) to data points are linear in $1/N^2$. To ensure better visibility of the fits at different $g$ values, $\ln g$ has been added.   The extrapolation to the continuum limit (symbol at infinite $N$) determines the coefficient $c/2$ of the divergent ($\sim N^2$) contribution in \eqref{fit}-\eqref{fit2} and is represented in the diagram of the right of this figure.    {\bf Right panel}: Data points estimate the continuum value of $c/2$ as from the extrapolations of the linear fits above. The simulations  at $g=30$, $m\,L=6$ (orange point) are used for a check of the finite volume effects, which appear here to be visible. Dashed and dotted lines are the results of, respectively, a linear fit in $1/g$ and a fit to a polynomial of degree two.}
     \label{fig:constant}
\end{figure}
\begin{figure}[h]
    \centering
     \includegraphics[scale=1]{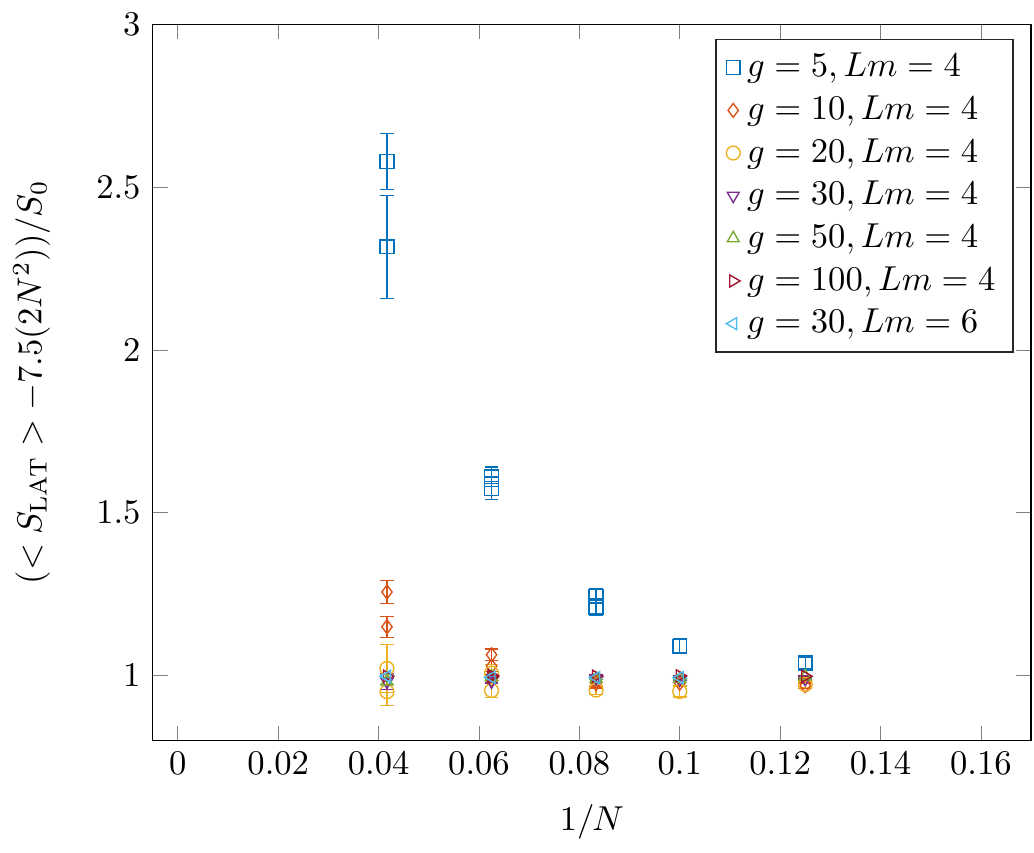}
        \caption{Plot of the ratio  $\frac{\langle S_{LAT}\rangle-\frac{c}{2}\, (2N^2)}{S_0} \equiv \frac{f'(g)}{4}$, where the coefficient of the divergent contribution $c$ has been here \emph{fixed} to the exact value  $c=15$ and $S_0=\frac{1}{2}M^2\,(2N^2)\,g$. For very large $g$, there is agreement with the continuum prediction $ f'(g) =4$ in  \eqref{cuspperturbative}.      For smaller values ($g=10, 5$, orange and light blue data points) strong deviations appear, compatible with quadratic divergences.  } 
     \label{fig:action_div}
\end{figure}
\begin{figure}[t]
    \centering
     \includegraphics[scale=1]{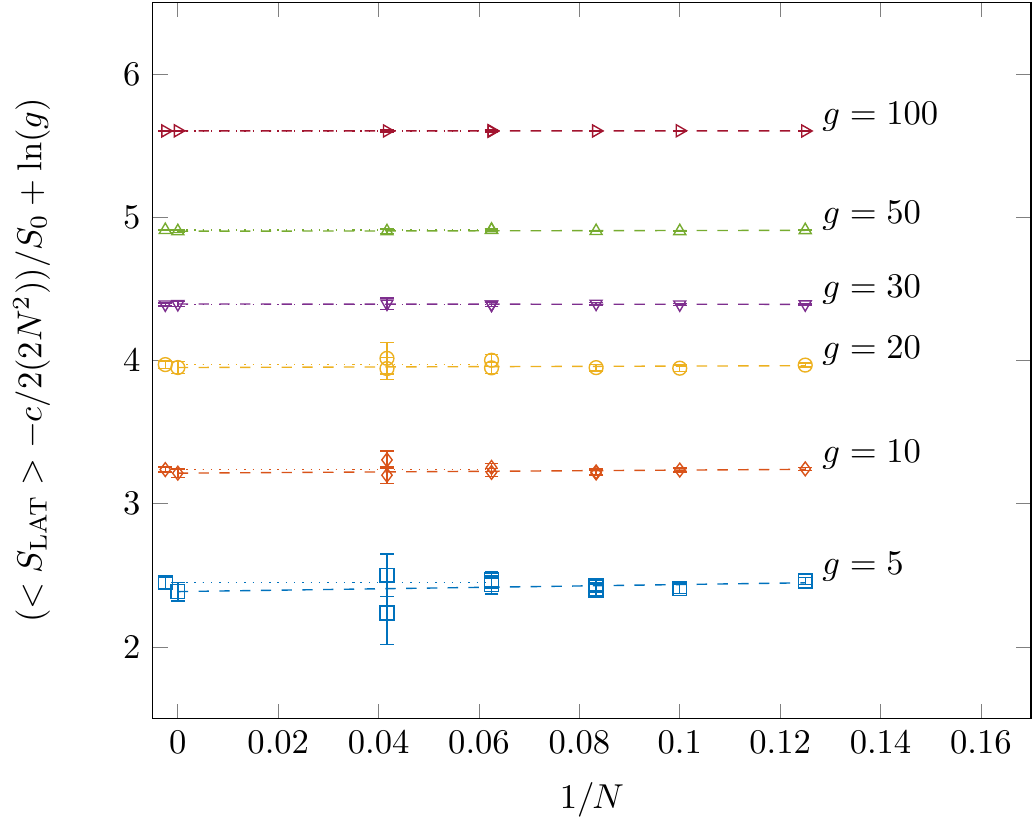}
          \caption{Plots for the ratio $\frac{\langle S_{LAT}\rangle-\frac{c}{2}\, (2N^2)}{S_0}+\ln g $ as a function of $1/N$, where the divergent contribution $c\,N^2/2$ is now the continuum extrapolation determined in Fig. \ref{fig:constant}. 
        To ensure better visibility of the fits at different $g$ values, $\ln g$ has been added.  
       Dashed lines represent a linear fit to all the data points for one value of $g$, while for dotted lines the fit is to a constant and only includes the two smallest lattice spacings.
        Symbols at zero (infinite $N$) are extrapolations from the fit constant in $1/N$.
           } 
     \label{fig:action_fin_N2}
\end{figure}
\begin{figure}[h]
    \centering
   \includegraphics[scale=1]{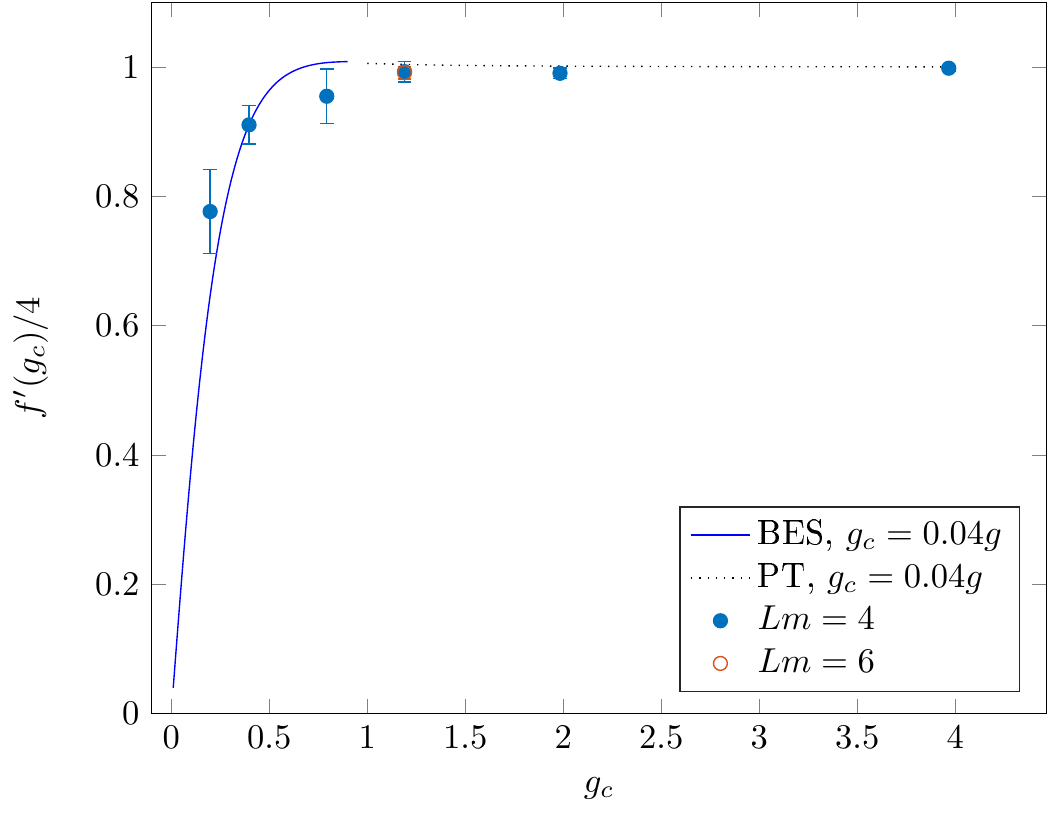}  
        \caption{Plot for $f'(g)/4$ as determined from the $N\to\infty$ extrapolation of \eqref{fit2}, i.e. from the extrapolations of the fits in Fig. \ref{fig:action_fin_N2}, and plotted as a function of the (bare) continuum coupling $g_c$ under the hypothesis that the latter is just a finite rescaling of the lattice bare coupling $g$ ($g_c=0.04\,g$), see discussion at the end of Section \ref{sec:simulation}.  The dashed line represents the first few terms in the perturbative series \eqref{cuspperturbative}, the continuous line is obtained from a numerical solution of the BES equation and represents therefore the prediction from  the integrability of the model. The simulations  at $g=30$, $m\,L=6$ (orange point) are used for a check of the finite volume effects, that appear to be within statistical errors.  } 
     \label{fig:action_fin_g}
\end{figure}
In measuring  the action \eqref{vevaction} on the lattice, exploring first the ``weak coupling'' (large $g$) region we are  supposed to recover the following general linear behavior in $g$~\footnote{We omit the label ``cusp'' in what follows.}
\be \label{fit}
\langle S_{\rm LAT}\rangle\equiv  \frac{c}{2}  (2N^2) +S_0\,,\qquad\qquad g\gg1, ~~~\qquad \text{where}~~S_0=\frac{1}{2}\,(2N^2)\,M^2\,g\,.
\ee
Above, we  reinserted the parameter $m$, used the leading, classical behavior $f(g)=4\,g$ in \eqref{cuspperturbative}, and used that $V_2\equiv T\,L=a^2\,(2N^2)$ since, as written above,  in our simulations the lattice temporal extent $T$ is always twice the spatial extent $L$ (therefore $T=a\,2N=2L$). We also introduced $S_0$ (which is \emph{linear} in $g$) for later convenience, to remind that in each simulation --  performed at fixed $g$ and at fixed $(N\,M)^2$ -- $S_0$ is also fixed. In \eqref{fit} we also added~$\frac{c}{2}  N^2$, namely a contribution constant in $g$ and (in the continuum limit $N\to\infty$) quadratically divergent.  This constant can be extrapolated for very large values of $g$ with a  fit  linear in $\frac{1}{N^2}$ from data points for $\frac{\langle S\rangle}{2N^2}=\frac{c}{2}+\frac{S_0}{2N^2}$. For $g=100, 50, 30$ this gives $c/2=7.5(1)$ -- red, green and violet fits in Fig. \ref{fig:constant}, left, respectively~\footnote{Recall that in Fig. \ref{fig:constant}  $\ln g$ has been added to ensure better visibility of the fits at different $g$ values.} -- consistently  with the number $15=8+7$ of bosonic fields appearing in the path integral. Namely, such a contribution to the  vev $\langle S\rangle=-\partial \ln Z/\partial \ln g$ in \eqref{fit}, field-independent and proportional to the lattice volume,  is simply counting the number of degrees of freedom which appear quadratically, and multiplying $g$, in the action.  Indeed, for very large $g$ the theory is quadratic in the bosons~\footnote{In lattice codes, it is conventional to omit the coupling form  the (pseudo)fermionic part of the action, since this is quadratic in the fields and hence its contribution in $g$ can be evaluated by a simple scaling argument.} 
 and equipartition holds, namely  integration over the bosonic variables yields a factor proportional to $g^{-\frac{(2N^2)}{2}}$ for each bosonic field species~\footnote{It is interesting to mention that in theories with exact supersymmetry this  constant contribution of the bosonic action (this time on the trivial vacuum) is valid at all orders in $g$, due to the 
 coupling constant independence of the free energy. For twisted
$\mathcal{N}=4$ SYM this is the origin of the supersymmetry Ward identity $S_{\rm bos}=9N^2/2$ per lattice site,  one of the observables used to measure soft supersymmetry breaking, see~\cite{Catterall:2008}. We thank David Schaich and Andreas Wipf for pointing this out to us.}.  

Having determined with good  precision the coefficient of the divergence, we can proceed first fixing it to be exactly $c=15$ and subtracting  from $\langle S_{\rm LAT}\rangle$ the corresponding contribution. Having in mind an analysis at finite $g$, we perform simulations in order to determine the ratio
\be \label{fit2}
\frac{\langle S_{LAT}\rangle-\frac{c}{2}\, (2N^2)}{S_0} \equiv \frac{f'(g)_{\rm LAT}}{4}\,.
\ee
On the right hand side we restored the general definition \eqref{vevaction}, which is the main aim of our study here.  
At $g=100, 50, 30, 20$ the plots in Fig.~\ref{fig:action_div} show a good agreement with the leading order prediction in  \eqref{cuspperturbative} for which $ f'(g) =4$.    For lower values of $g$ -- orange and light blue data points in Figure \ref{fig:action_div}  -- we observe deviations that obstruct the continuum limit and signal  the presence of further quadratic ($\sim N^2$) divergences.  They are compatible with an Ansatz for $\langle S_{\rm LAT}\rangle$ for which the ``constant'' contribution multiplying $2N^2$ in \eqref{fit}-\eqref{fit2} is actually  $g$-dependent.  
It seems natural  to relate these power-divergences to those arising in continuum perturbation theory, where they are                                                            usually set to zero using dimensional regularization~\cite{Giombi}. From the perspective of a hard cut-off regularization like the lattice one, this is related to the emergence in the continuum limit of power divergences -- quadratic, in the present two-dimensional case -- induced by mixing of the (scalar) Lagrangian with the identity operator under UV renormalization.   Additional contributions to these deviations might be due to the (possibly wrong) way the continuum limit is taken, \emph{i.e.} they could be related to a  possible infinite renormalization occurring in those field correlators and corresponding physical masses which have been  not investigated here (fermionic and $z$ excitations). 
%, and to the additional mixing cause by a Wilson term explicitly breaking  the original $SO(6)$ symmetry of  \eqref{S_cusp}.  
While to shed light on the issue  such points should be investigated in the future -- see further comments in Section \ref{sec:conclusions} --  we proceed  with a non-perturbative subtraction of these divergences. Namely, from the data of Fig. \ref{fig:action_div} we subtract the continuum extrapolation of $\frac{c}{2}$ (multiplied by the number of lattice points, $2N^2$), as determined in the right diagram of Fig. \ref{fig:constant},  for the full range of the coupling explored. The result is shown in Fig. \ref{fig:action_fin_N2}.  The divergences appear to be completely subtracted, confirming their purely quadratic nature.
The flatness of data points - which can be fitted by a constant -- indicates very small lattice artifacts. At least in the region of lattice spacings explored from our simulations errors are small, and do not diverge as one approaches the $N\to\infty$ limit. 
We can thus use the  extrapolations at infinite $N$ of Fig. \ref{fig:action_fin_N2}  to show  the continuum limit for the left hand side of \eqref{fit2}, Fig. \ref{fig:action_fin_g}. This is our measure for $f'(g)/4$, and it allows in principle a direct comparison with the perturbative series (dashed line) and with prediction obtained via the integrability of the model (continuous line, representing the first derivative of the cusp as obtained from a numerical solution of the BES equation~\cite{BES}~\footnote{We thank D. Volin for providing us with a numerical solution to the BES equation.}). 
To compare our extrapolations with the continuum expectation, we  match the lattice point for the observable $f'(g)$ at $g=10$ -- as determined from the~$N\to\infty$ limit of $f'(g)_{\rm LAT}$ \eqref{fit2} -- with the continuum value for the observable $f'(g_c)_c$ as determined from the integrability prediction, \emph{i.e.} as obtained from a numerical solution of the BES equation~\cite{BES}. This is where in Fig. \ref{fig:action_fin_g} the lattice point lies exactly on the (integrability) continuum curve.  
The value $g=10$ has been chosen as a reference point since it is far enough from both the region where the observable is substantially flat and proportional to one (which ensure a better matching procedure)
and the region of higher errors (also, where the sign problem plays no role yet, see Section \ref{sec:phase}). %in contrast with s far enough from the larger in the corresponding  and because it 
Assuming that a simple finite rescaling relates the lattice bare coupling $g$ and the (bare) continuum one $g_c$,  from $f'(g)= f'(g_c)_c$ we then derive  that  $g_c=0.04 g$. 
 A simple look at  Fig. \ref{fig:action_fin_g} shows that, in the perturbative region, our analysis -- and the related assumption for the finite rescaling of the coupling -- is in good qualitative agreement with the integrability prediction. 
About direct comparison with the perturbative series \eqref{cuspperturbative}, since we are considering the  derivative of \eqref{cuspperturbative}  the first correction to the expected large $g$ behavior $f'(g)/4\sim 1$  is  positive and proportional to the Catalan constant $K$. The plot in Fig. \ref{fig:action_fin_g} does not catch the upward trend of such a  first correction (which is too small, about $2$ percent, if compared to the statistical error). % but only the bending down due to higher order corrections.  
Notice that, again under the assumption that such simple relation between the couplings exists  -- something that within our error bars cannot be excluded  --  the nonperturbative regime beginning with $g_c=1$ would start at $g=25$, implying that our simulations at $g=10, 5$  would  already test a fully non-perturbative regime of the string sigma-model under investigation.   The mild discrepancy observed in that point of this region  ($g=5$ or $g_c=0.2$) which is not fixed by definition via the ``matching'' procedure discussed above could be the effect of several contributing causes. Among them, systematic factors as the ones related to the complex phase -- and its omission from the measurements, see below --  as well as finite volume effects with related errors in the non-perturbative subtraction of divergences. We  emphasize that  the relation between the lattice and continuum bare couplings might well be \emph{not} just a finite rescaling. 
To shed light on this point, the  matching procedure should  use points at further smaller values of $g$. 
We summarize and further comment these questions in Section~\ref{sec:conclusions}, and discuss in more detail one of the most relevant issues~--~the observed complex phase which inhibits measurements at the interesting, small values of $g$~--~in the next section. 
\subsection{The phase}
\label{sec:phase}

 \begin{figure}[t]
   \centering
 \includegraphics[scale=0.7]{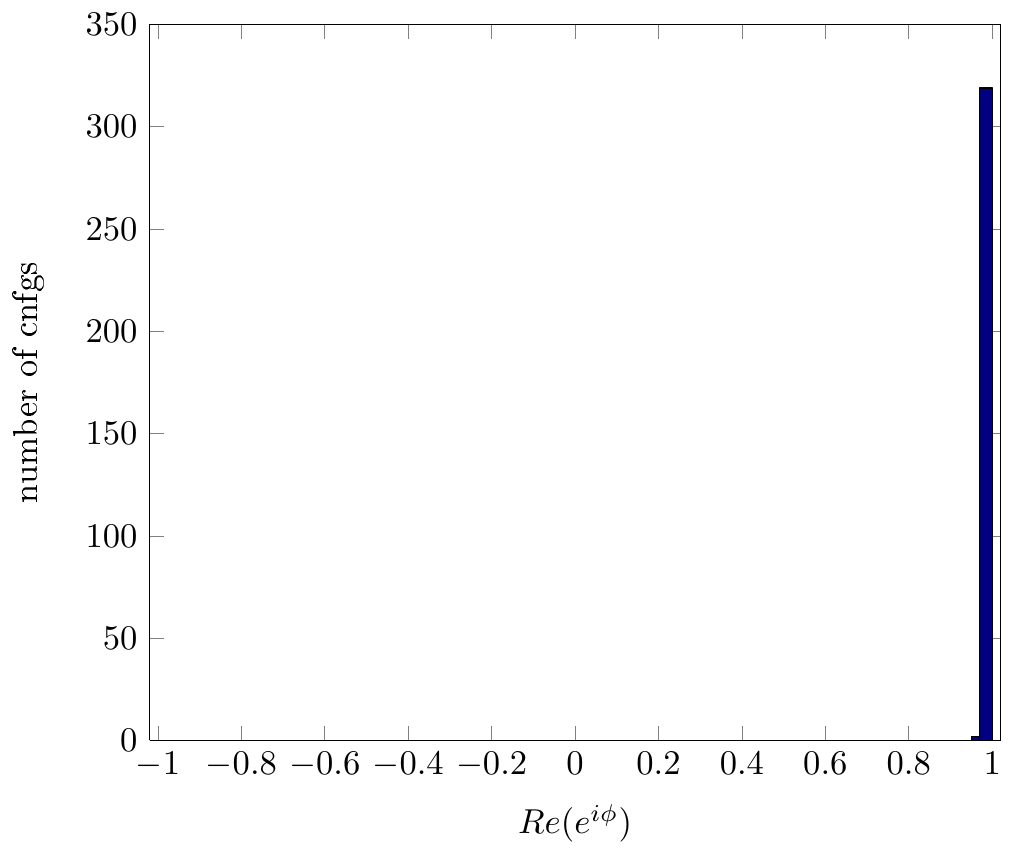}
  \includegraphics[scale=0.7]{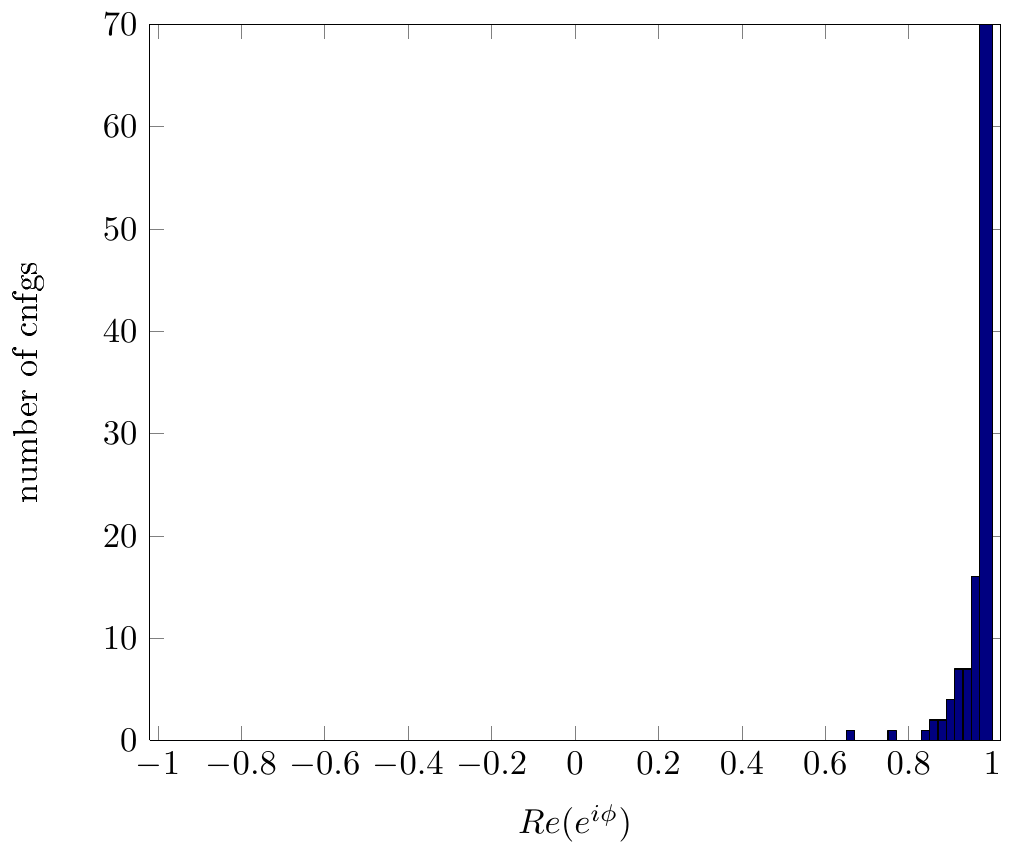}
   \includegraphics[scale=0.7]{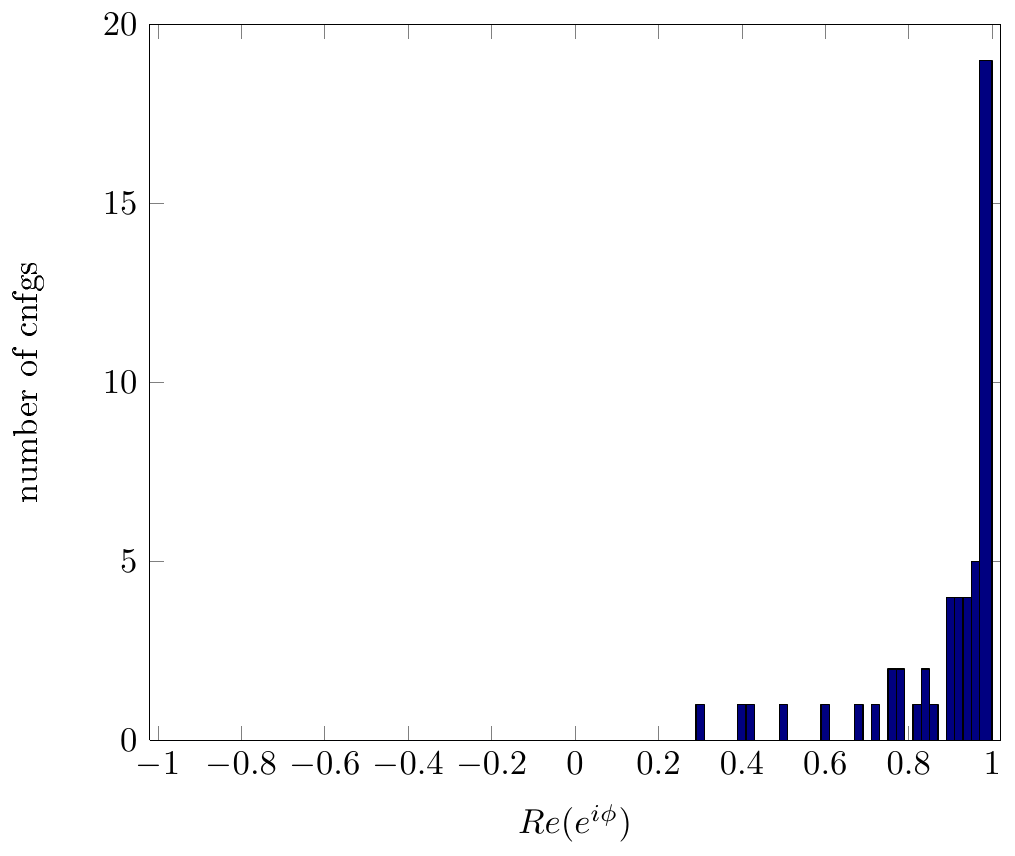}
    \includegraphics[scale=0.7]{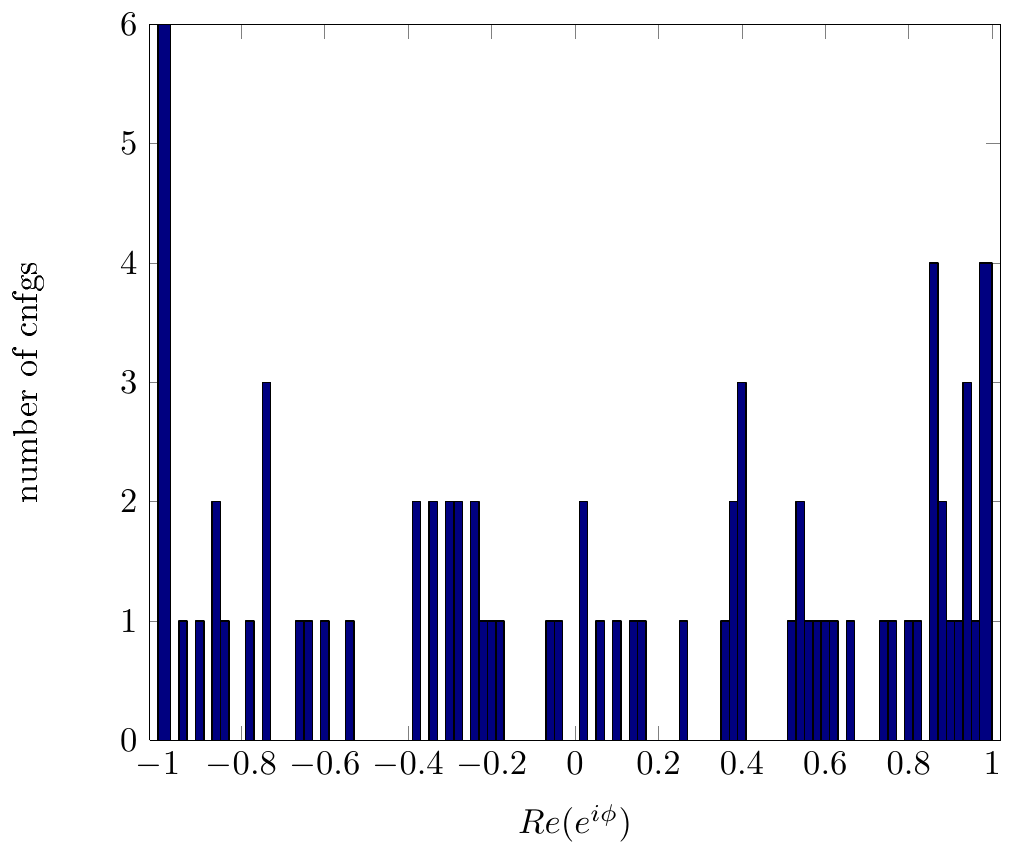}
 \caption{Histograms for the  frequency of the real part %it vanishes by symmetry, and also it must be zero
 of the   reweighting phase factor $e^{i\theta}$ of the Pfaffian ${\rm Pf}\,O_F=|(\det O_F)^{\frac{1}{2}}|\,e^{i\theta}$, based on the ensembles generated  at  $g=30,10,5,1$ (from left to right, top to down) for $L/a=8$. The plots here shown use the discretization \eqref{OFgenbreak}-\eqref{Wilsonshiftgenbreak}, however we found no substantial difference between this analysis and the one performed with the discretization \eqref{OFgen}-\eqref{Wilsonshiftgen}.}
\label{fig:phase}
\end{figure} 

 After the linearization realized via the Hubbard-Stratonovich transformation  \eqref{HubbardStratonovich}, the formal integration over the fermionic components leads to a Pfaffian.  For any given bosonic configuration, the latter is manifestly \emph{not} real. As discussed in Section \ref{sec:continuum}, one of the Yukawa terms resulting from  linearization -- specifically, this is the last term in the second line of \eqref{HubbardStratonovich} -- introduces a phase, so that 
${\rm Pf}\,O_F=|(\det O_F)^{\frac{1}{2}}|\,e^{i\theta}$. The standard way to proceed is to perform ``phase-quenched'' simulations, omitting $e^{i\theta}$ from the integration measure which includes only the absolute value of the Pfaffian, employing pseudofermions as in  \eqref{fermionsintegration}. 
Such a procedure ensures drastic computational simplifications, and  still can deliver the true expectation value of the observable under analysis via  phase reweighting. Namely,  the nonpositive part of the Boltzmann weight (which is the complex phase) is incorporated into the observable in the measurement 
\be\label{reweight}
\langle S  \rangle_{\rm reweight}=\frac{\langle S\, e^{i\theta}\rangle}{\langle e^{i\theta}\rangle}~.
\ee
%With a positive real  Pfaffian at non-zero lattice spacing, one would have  $\langle e^{i\theta}\rangle=1$, and 
%\colb{If $\langle e^{i \theta}\rangle$ is finite (non-zero) and in absence of correlation between the factors in the numerator of \eqref{reweight}} it is
%$\langle S  \rangle_{\rm reweight}= \langle S\,  \rangle$. 
%
If $\langle e^{i \theta}\rangle$ averages to  zero (due to $\theta$ fluctuating far from zero on a significant part  of the bosonic field configurations generated via phase-quenched approximation) the reweighting procedure breaks down. The corresponding sign problem is known to be a serious   obstacle for numerical simulations. 

We have explicitly computed the reweighting (phase) factor for smaller lattices, up to~$L/a=12$, and observed that the reweighting has no effect on the central value of the  two observables that we study -- namely, for the observables $\mathcal{O}$  it holds $\langle \mathcal{O}  \rangle_{\rm reweight}= \langle \mathcal{O}\,  \rangle$ within errors~\footnote{This is suggesting the absence of correlation between the two factors in the numerator of \eqref{reweight}.}.  
Thus, in the analysis presented in the previous sections and in Appendix \ref{app:altern_discretization} we omit the phase from the simulations in order to be able to consistently take the continuum limit. In absence of data for the phase factor in the case of larger lattices, we do not assess the possible systematic error related to this procedure. 

To explore the possibility of a sign problem in simulations, we have then studied the relative frequency for  the real part (the imaginary part is zero within errors, as predicted from the reality of the observables studied) of the  Pfaffian phase $e^{i\theta}$, as shown in Figure \ref{fig:phase},  at $g=30,10,5,1$ (from left to right, top to down). 
At $g=1$, right bottom histogram, the observed $\langle e^{i\theta}\rangle$ is consistent with zero, thus preventing the use of standard reweighting. In the sense explained above, the analysis we present here is thus also limited to the values $g=100,50,30,20,10,5$ of the coupling (and with the further parameters listed in Tables \ref{t:runs} and \ref{t:runs2}).
Therefore, a severe sign problem is appearing precisely for values of the coupling referring to a fully non-perturbative regime (corresponding to weakly-coupled $\mathcal{N}=4$ SYM). Therefore, in order to investigate this interesting and crucial region of the couplings alternative algorithms or settings (in terms of a different, phase-free linearization) should be considered.

\section{Conclusions}
\label{sec:conclusions}

In this paper we have considered two possible discretizations for the AdS-lightcone gauge-fixed  action for the Type IIB Green-Schwarz  superstring. We have used them  for measuring the (derivative of the) cusp anomalous dimension of planar $\mathcal{N}=4$ SYM as derived from string theory, as well as the masses of  two bosonic fields, namely the $AdS$ Lagrangian excitations  transverse to the relevant, classical string solution.    In both cases, our continuum extrapolations show a good agreement (qualitative for the mass and quantitative for the action) in the large $g=\sqrt{\lambda}/(4\pi)$ regime, which is the perturbative regime of the sigma-model. For smaller values of $g$,  further work appears to be necessary to address both numerical and conceptual challenges indicated by our analysis.

Lattice simulations were performed  employing  a Rational Hybrid Monte Carlo (RHMC) algorithm and two Wilson-like fermion discretizations, breaking different subgroups of the global symmetry for the relevant sigma-model. 
Interestingly, continuum results seem not to be sensitive to the differences in the discretisation. Our line of constant physics demands physical masses to be kept constant while approaching the continuum limit, which in the case of finite mass renormalization requires no tuning of the ``bare'' mass parameter of the theory (the light-cone momentum $P^+$). For one of the bosonic fields entering the Lagrangian we determine the correlator and physical mass, confirming the expected finite renormalization and thus no need of tuning. At large $g$, the mentioned agreement of our measures with the expected behavior for both the mass and the action is very encouraging. 

In measuring the action at small values of the coupling $g$, we observe a divergence compatible with a quadratic behavior $\sim a^{-2}$. 
It is certainly possible that the reasoning leading to  the line of constant physics \eqref{constantphysics} %motivated by the study of  just one species of correlators, 
might be subject to change once all fields correlators are investigated -- something which we leave for the future. However, in the lattice regularization performed here such divergences are expected.  In continuum perturbation theory, power-divergences arising in this~\cite{Giombi} and analogue models~\cite{abjm2loops}  are set to zero using dimensional regularization. From the perspective of a hard cut-off regularization like the lattice one, this is related to the emergence in the continuum limit of power divergences -- quadratic, in the present two-dimensional case -- induced by mixing of the (scalar) Lagrangian with the identity operator under UV renormalization.  The problem of renormalization in presence of power divergences is in general non trivial, and one of the ways to proceed -- which is our way here -- is via non-perturbative subtractions of those divergences. 
While with the present data we are able to reliably and non-perturbatively subtract them, 
in general this procedure leads   to potentially severe ambiguities, with errors diverging in the continuum limit. 
In the future it may be therefore worthwhile to explore whether other schemes -- e.g.  the Schr\"odinger functional scheme~\cite{Luscher:1992an} -- could be used as a proper definition of the effective action  under investigation. 
%redefinitions of the effective action under investigation 
%whether other schemes -- e.g. employing  the Schr\"odinger functional scheme~\cite{Luscher:1991wu}. 
We remark however that for the other physical observable here investigated, the $\langle x\,x^*\rangle$ correlator, we encountered no problems in proceeding to the continuum limit. 

For both observables, the comparison of our continuum extrapolation with the predictions coming from integrability -- Figs. \ref{fig:correlator} and \ref{fig:action_fin_g} -- is done matching at a given coupling the corresponding values for the continuum extrapolation of $f'(g)_{\rm LAT}$ and the integrability prediction $f'(g_c)_c$. Assuming that a simple finite rescaling relates the lattice bare coupling $g$ and the (bare) continuum one $g_c$,  one simply derives that  $g_c=0.04 g$ and proceeds with the comparison of further data points.  It might well be that this assumption is wrong, which could be supported from further data at smaller values of $g$ -- something  at present inhibited by the sign problem occurring there --  and would also explain the (mild) discrepancy observed in Fig. \ref{fig:action_fin_g} at $g=5$. Clearly,  a non-trivial relation between $g$ and $g_c$  would take away any predictivity from the lattice measurements for the (derivative of the) cusp. To proceed, one could then 
define the continuum (BES) prediction as the point where to study the theory and tune accordingly the lattice bare coupling, \emph{i.e.} numerically determine such non-trivial interpolating function of the bare couplings. This could then be used as an input for the~-- this time fully predictive~--~measurements of other physical observables (like the mass $m^2_x$ here).

As mentioned, our results seems not to be sensitive to the discretization adopted. 
We used a discretization which breaks an $SO(2)$ rotational symmetry in the two $AdS_5$ directions orthogonal to $AdS_3$ (i.e. transverse to the classical solution), and analized the observables also in another setting (see Appendix~\ref{app:altern_discretization}) where the Wilson-like term explicitly breaks the $SO(6)$ symmetry of the model.  Since both the observables we study -- $f(g)$ and $x,x^*$ correlators -- are  $SO(6)$ singlets, we would expect  significant differences only in the way the continuum limit is taken (mainly due to the larger mixing pattern in the UV renormalization for simulations with broken $SO(6)$ symmetry). However, at least in the range of the coupling explored, this does not seem to be the case. Furthermore, the continuum extrapolations of the same observable in the two different discretizations agree within errors, which is strongly suggesting that the two discretizations lead to the same continuum limit.

%There are still conceptual and numerical issues that remain to be clarified before one would be able to state that,  even only for the example of the cusp anomaly, the discretization and the continuum limit proposed in this paper are ÒconsistentÓ and might be exported to the study of further observables. 

One further important result of our analysis is the detection of a phase in the fermionic determinant, resulting from integrating out the fermions. This phase  is caused by the linearization of fermionic interactions introduced in \cite{Roiban}. 
%resulting from integrating out the fermions, which is introduced by the linearization of fermionic interactions used in \cite{Roiban}. 
 %, which is however very small with respect to the one implicit in the Hubbard-Stratonovich transformation.}. 
%Standard reweighting works in the perturbative sigma-model regime,
For values of the coupling approaching the non-perturbative regime (corresponding to weakly-coupled $\mathcal{N}=4$ SYM)  the phase undergoes strong fluctuations, signaling a severe sign problem. 
It would be desirable to find alternative ways to linearize quartic fermionic interactions, with resulting Yukawa terms leading to a real, positive definite fermionic determinant. 
%This would allow measurements for smaller values of $g$, crucial to shed light on the 
%

Progress about these issues  is ongoing and we hope to report on it in the near future. 
%We hope to be able to report progress about these issues in the near future.  

\section*{Acknowledgements}

We are particularly grateful to M. Bruno for initial collaboration, and to R. Roiban, D. Schaich, R. Sommer, S. Sint and A. Wipf  for very useful elucidations. We thank D. Volin for providing us with a numerical solution to the BES equation. We also thank F. Di Renzo, H. Dorn, S. Frolov, G. Eruzzi, M. Hanada and the theory group at Yukawa Institute, B. Hoare, B. Lucini, T. Mc Loughlin, J. Plefka, A. Schwimmer, S.  Theisen, P. T\"opfer, A. Tseytlin and U. Wolff for useful discussions.
 
 %%%%%%%%%%%%%%%%%%%%%%%%%%%%%%%%%%%%%
 %%%%%%%%%%%%%%%%%%%%%%%%%%%%%%%%%%%%%

\appendix

\section{The model in the continuum}
\label{app:continuum}

In this Appendix we briefly recall the steps leading to the action \eqref{S_cusp}.
 
The $AdS_5\times S^5$ background metric in Poincar\'e patch is (setting to 1 the radius of both $AdS_5$ and $S^5$)
\begin{eqnarray}\nonumber
&&ds^2=z^{-2}\,(dx^m\,dx_m+dz^M\,dz^M)=z^{-2}(dx^m\,dx_m+dz^2)+du^M du^M\\\label{adsmetric}
&&x^m x_m=x^+ x^-+x^* x\,,\qquad x^\pm=x^3\pm x^0\,,\qquad x=x^1+i x^2\,,\\\nonumber
&&z^M=z\,u^M\,, \qquad u^M\,u^M=1\,\qquad z=(z^M z^M)^{\frac{1}{2}}~.
\end{eqnarray}
Above, $x^\pm$ are the light-cone coordinates, $x^m=(x^0,x^1,x^2, x^3)$ parametrize the 
four-dimensional  boundary of $AdS_5$ and $z\equiv e^{\phi}$ is the radial coordinate.

The  AdS light-cone gauge~\cite{MT2000,MTT2000} is defined by fixing the local symmetries of the superstring action, bosonic diffeomorphisms and $\kappa$-symmetry, via a sort of ``non-conformal'' gauge and a more standard light-cone gauge   on the two  Majorana-Weyl  fermions of Type IIB superstring  action respectively  
 as follows~\footnote{As in the standard conformal gauge,  the choice $x^+ = p^+ \tau$ is allowed by residual
diffeomorphisms after the choice \eqref{bosgauge}.}
\begin{eqnarray}  \label{bosgauge}
&&  \sqrt{-g} g^{\a\b} = {\rm diag}(-z^2, z^{-2})\ , \qquad \qquad x^+ = p^+ \tau \ ,\\
&&\Gamma^+ \theta^I=0
\end{eqnarray}
  
The  resulting \adss   superstring  action  can be written as ($z^M=z \, u^M$)
\begin{eqnarray}
\label{s}
S &=& \frac{1}{2} T \int d \tau \int 
 d \sigma \; \mathcal{L}\ , \quad  \quad  \quad T =
\frac{R^2}{2 \pi \alpha'} = \frac {\sqrt{\lambda}}{2 \pi} \ , \\
\mathcal{L} &=& \dot{x}^* \dot{x} + (\dot z^M  + \mathrm{i}  p^+ z^{-2} z^N 
\eta_i {\rho^{MN}}^i{}_j \eta^j)^2  + \mathrm{i} p^+ (\theta^i \dot{\theta}_i +
       \eta^i\dot{\eta}_i +\theta_{i}\dot{\theta}^{i}+\eta_{i}\dot{\eta}^{i}
)+\no \\
&&
  \quad  - (p^+)^2 z^{-2} (\eta^2)^2 - z^{-4} ( x'^*x'  + {z'}^M {z'}^M) 
  \no \\
&&
  \quad - 2 \Big[\ p^+ 
       z^{-3}\eta^i \rho_{ij}^M z^M (\theta'^j - \mathrm{i}
       z^{-1} \eta^j  x') +  p^+ 
       z^{-3}\eta^i (\rho^\dagger_M)^{ij} z^M (\theta'^j + \mathrm{i}
       z^{-1} \eta^j  x'^*\Big]\; \label{la}\\
       &\equiv&\dot{x}^* \dot{x} + (\dot z^M  + \mathrm{i}  p^+ z^{-2} z^N 
\eta_i {\rho^{MN}}^i{}_j \eta^j)^2  + \mathrm{i} p^+ (\theta^i \dot{\theta}_i +
       \eta^i\dot{\eta}_i - h.c.) - (p^+)^2 z^{-2} (\eta^2)^2 \no \\
&&
  \quad - z^{-4} ( x'^*x'  + {z'}^M {z'}^M) - 2 \Big[\ p^+ 
       z^{-3}\eta^i \rho_{ij}^M z^M (\theta'^j - \mathrm{i}
       z^{-1} \eta^j  x') + h.c.\Big]\;.  \label{la}
\end{eqnarray}
%The action \eqref{s}  is \emph{real}. 
%~\footnote{For the terms in the round and square brackets one should take into account, respectively, that
%\begin{eqnarray}
%&&\Big( \mathrm{i}  \eta_i {\rho^{MN}}^i{}_j \eta^j\Big)^\dagger=-\mathrm{i}\,(\eta^j)^\dagger({\rho^{MN}}^i{}_j)^*(\eta_i)^\dagger
%=-\mathrm{i}\,\eta_j\,{\rho^{MN}}_i{}^j\,\eta^i=+\mathrm{i}\,\eta_j\,{\rho^{MN}}^j{}_i\,\eta^i\equiv   \mathrm{i}  \eta_i {\rho^{MN}}^i{}_j \eta^j
%\\
%&&\Big(\eta^i(\rho^M)_{ij}\theta'^j\Big)^\dagger=(\th'^j)^\dagger(\rho^M_{ij})^*(\eta^i)^\dagger=-\th'_j(\rho^M)^{ij}\eta_i=\eta_i(\rho^M)^{ij}\theta'_j~.
%\end{eqnarray}
 %}, and 
 % enters the partition function as $Z=e^{i\,S}$. 
 Wick-rotating $\tau \to  -\mathrm{i} \tau, \ p^+ \to  \mathrm{i} p^+$, and setting $p^+=1$, one gets $Z=e^{-S_E}$, where $S_E = \frac{1}{2} T \int d \tau d \sigma \; \mathcal{L}_E$  and 
%%%%%%%%%%%%%%%%%%%%%%%%%%%%%%%%%%%%%%%%%%%%%
\begin{flalign}\nonumber
\mathcal{L}_{E} & =\dot{x}^{*}\dot{x}+ \big(\dot{z}^{M}+i\,z^{-2}z_{N}\eta_{i}(\rho^{MN})^i_{\hphantom{i} j}\eta^{j}\big)^{2}+i\big(\theta^{i}\dot{\theta}_{i}+\eta^{i}\dot{\eta}_{i}-\text{h.c.}\big)-z^{-2}\left(\eta^2\right)^{2}\\\label{LCADS_euc}
&+z^{-4} (x^{'*}x^{'}+z^{'M}z^{'M})
 +2i\Big[z^{-3}z^{M}\eta^{i}{\rho^{M}}_{ij}\big(\theta^{'j}-i\,z^{-1}\eta^{j}x^{'}\big)+\text{h.c.}\Big]
 \end{flalign}
%\footnote{The l.c. gauge-fixed \emph{euclidean} action is not real because of the term in square brackets, since $(i[...+\text{h.c.}])^\dagger=-i (...+\text{h.c.})$. 

%%%%%%%
\noindent
The null cusp background
\begin{gather}
\label{null_cusp_back}
x^{+}=\tau \qquad \qquad x^{-}=-\frac{1}{2 \sigma} \qquad \qquad x=x^{*}=0 \qquad \qquad z=\sqrt{\frac{\tau}{\sigma}}\,,\qquad\qquad  \tau,\sigma>0\,,
\end{gather}
is the classical solution of the string action that describes a Euclidean open string surface ending on a lightlike Wilson cusp in the AdS boundary at $z=0$ \cite{Giombi}. This string vacuum is actually degenerate as any $SO(6)$ transformation on $z^M$ leaves the last condition above unaltered. The fluctuation spectrum of this solution can be easily found by fixing a direction, say $u^M=(0,\,0,\,0,\,0,\,0,\,1)$, and defining the fluctuation fields
\begin{eqnarray}
&& z=\sqrt{\frac{\tau}{\sigma}}\ {\tilde z} \ , \ \ \ \ \ \ \ \ 
{\tilde z} = e^{\tilde \phi}= 1 + \tilde \phi  +\dots~,\ \ \  
 z^M=\sqrt{\frac{\tau}{\sigma}}\ {\tilde z}^M \ , \ \ \ \ 
{\tilde z}^M = e^{\tilde \phi} \tilde u^M  \nonumber\\
&&
{\tilde u}{}^{a}=  \frac{y^{a}}{1+\frac{1}{4}y^2}~, \ \ \ \ 
{\tilde u}{}^{6} =  \frac{1-\frac{1}{4}y^2}{1+\frac{1}{4}y^2}  \ , \ \ \ \ \ \ \ \ \
~~~~ y^2\equiv \sum_{a=1}^5 (y^a)^2\ , \ \ \ \ \ a=1,...,5 \ , \label{exp6}\\\nonumber
&&
x = \sqrt{\frac{\tau}{\sigma}} \ {\tilde x}
~,~~~~~~
\theta=\frac{1}{\sqrt{\sigma}}{\tilde\theta}
~,~~~~~~
\eta=\frac{1}{\sqrt{\sigma}}{\tilde\eta} \,. 
\end{eqnarray}
The further redefinition of the worldsheet coordinates
\begin{gather}
t=\log\tau \qquad \qquad s=\log\sigma\,
\end{gather}
which absorb powers of $\tau,\, \sigma$ so that the resulting fluctuation Lagrangian has constant coefficients, leads to 
the Lagrangian $\mathcal{L}_{\rm cusp}$ in \eqref{S_cusp}. 
If we truncate it at quadratic order in the fluctuations fields
\begin{eqnarray}
{\cal L}_2  &=&(\partial_t {\tilde \phi})^2+(\partial_s {\tilde \phi})^2 +{\tilde \phi}^2 + 
|\partial_t {\tilde x}|^2
     +|\partial_s {\tilde x}|^2
+\frac{1}{2}|{\tilde x}|^2 
+(\partial_t {y}^a)^2+(\partial_s {y}^a)^2\no
 \\\label{quadraticlagr}
&+&  2{\rm i}\;
({\tilde \theta}^i \partial_t{\tilde \theta}_i
+{\tilde \eta}^i\partial_t{\tilde \eta}_i
)
+2{\rm i}\; {\tilde \eta}^i (\rho^6)_{ij}
        (\partial_s{\tilde \theta}^j -{\tilde \theta}^j)
+
2{\rm i}\; {\tilde \eta}_i (\rho^\dagger_6)^{ij}
(\partial_s {\tilde \theta}_j - {\tilde \theta}_j) ~,
\end{eqnarray}
it is easy to see that the bosonic excitation spectrum consists of one field ($\tilde{\phi}$) with $m^2=1$, two fields ($x,x^*$) with $m^2=\frac{1}{2}$ and five fields $(y^a)$ with $m^2=0$ \cite{Giombi}. Adding the fermionic determinant as in \eqref{freefermdet}, this means that the full one loop effective action $\Gamma^{(1)}=-\ln Z^{(1)}$ reads
\be\label{oneloopcontinuum}
\Gamma^{(1)}=V_2  \frac{1}{2}\int \frac{dp_0 dp_1}{(2\pi)^2}\ln\Big[\frac{(p_0^2+p_1^2+1)(p_0^2+p_1^2+\frac{1}{2})^2(p_0^2+p_1^2)^5}{(p_0^2 + p_1^2 + \frac{1}{4})^8}\Big]=-\frac{3\ln 2}{8 \pi}\,V_2 \,.
\ee

%%%%%%%%%%%%%%%%%%%%%%%%%%%%%%%%%
%\bigskip
 
%%%%%%%%%%%%%%%%%%%%%%%%%%%%%%%%%%%%%%%%%%%%%

%For completeness, we  also report here the expression for the complex-conjugate of the quadratic fermionic operator obtained via the Hubbard-Stratonovich transformation \eqref{HubbardStratonovich}
%\begin{flalign}
%\!\!\!\!\!\!\!
% O_F^\dagger & =\left(\begin{array}{cccc}
%0 & i\partial_{t} & \mathrm{i}\rho^\dagger_{M}\left(\partial_{s}+\frac{1}{2}\right)\frac{{z}^{M}}{{z}^{3}} & 0\\
%\mathrm{i}\partial_{t} & 0 & 0 &\mathrm{i}\rho^{M}\left(\partial_{s}+\frac{1}{2}\right)\frac{{z}^{M}}{{z}^{3}}\\
%-\mathrm{i}\frac{{z}^{M}}{{z}^{3}}\rho^\dagger_{M}\left(\partial_{s}-\frac{1}{2}\right) & 0 & 2\frac{{z}^{M}}{{z}^{4}}\rho \dagger_{M}\left(\partial_{s}{x}^*-\frac{{x}^*}{2}\right) & i\partial_{t}+A\\
%0 &- \mathrm{i}\frac{{z}^{M}}{{z}^{3}}\rho^{M}\left(\partial_{s}-\frac{1}{2}\right) &\mathrm{i}\partial_{t}-A^\dagger & -2\frac{{z}^{M}}{{z}^{4}}\rho^{M}\left(\partial_{s}{x}-\frac{{x}}{2}\right)
%\end{array}\right)
%\end{flalign}
%%where the operator $A$ is defined  in \eqref{Aoperator}.  
%%%%%%%%%%%%%%%%%%%%%%%%%%%%%%%%%%%%%%%%%%%%%
For the $SO(6)$ generators built out of the $\rho^{M}_{ij} $ of $SO(6)$ Dirac matrices it holds
\begin{equation}\label{rhomatrices}
\begin{split}
  (\rho^{MN})^i_{\hphantom{i}j}&=\frac{1}{2}({\rho^M}^{i\ell}\,\rho^N_{\ell j} -{\rho^N}^{i\ell}\,\rho^M_{\ell j} )=\frac{1}{2}(\rho^M_{i\ell}\,{\rho^N}^{\ell j} -\rho^N_{i\ell}\,{\rho^M}^{\ell j} )^*\equiv\left( (\rho^{MN})_i^{\hphantom{i}j}\right)^*\\  (\rho^{MN})^i_{\hphantom{i}j}&=-(\rho^{MN})_j^{\hphantom{j}i} \,\qquad\qquad  (\rho^{MN})_i^{\hphantom{i}j}=- (\rho^{MN})^j_{\hphantom{j}i}\,,
  \end{split}
\end{equation}
where in the last equation we used that $\frac{1}{2}({\rho^M}^{i\ell}\,\rho^N_{\ell j} -{\rho^N}^{i\ell}\,\rho^M_{\ell j} )=-\frac{1}{2}(\rho^M_{j\ell}\,{\rho^N}^{\ell i} -\rho^N_{j\ell}\,{\rho^M}^{\ell i} )$.
%Then
%\begin{eqnarray}
%\Big(\eta^i (\rho^{MN})_i^{\hphantom{i}j}\theta_j\Big)^\dagger&=&\frac{1}{2}(\theta_j)^\dagger\,[\rho^M_{i\ell}{\rho^N}^{\ell j}-{\rho^N}^{i\ell}\rho^M_{\ell j}]^*(\eta^i)^\dagger=\theta^j(\rho^{MN})^{i}_{\hphantom{i}j}\eta_i\\\nonumber
%&=&-\eta_i \,(\rho^{MN})^{i}_{\hphantom{i}j}\,\theta^j~.
%\ee
Useful flipping rules are
\begin{eqnarray}
\eta\,\rho^M\,\theta&=&\eta^i\,\rho^M_{ij}\,\theta^j=-\theta^j\,\rho^M_{ij}\,\eta^i=\theta^j\,\rho^M_{ji}\,\eta^i\equiv \theta^i\,\rho^M_{ij}\,\eta^j=\theta\,\rho^M\,\eta\\
\eta^\dagger\rho^\dagger_M\,\theta^\dagger&=&\eta_i\,{\rho^M}^{ij}\,\theta_j=-\theta_j\,{\rho^M}^{ij}\,\eta_i=\theta_j\,{\rho^M}^{ji}\,\eta_i\equiv \theta_i\,{\rho^M}^{ij}\,\eta_j=\theta^\dagger \rho^\dagger_M\,\eta^\dagger\\
\eta_i\,(\rho^{MN})^i_{\hphantom{i}j}\,\theta^j&=&-\theta^j\,(\rho^{MN})^i_{\hphantom{i}j}\,\eta_i=\theta^j\,(\rho^{MN})_j^{\hphantom{j}i}\,\eta_i\equiv\theta^i\,(\rho^{MN})_i^{\hphantom{i}j}\,\eta_j~.
\end{eqnarray}

\section{Alternative discretization}
\label{app:altern_discretization}

\begin{figure}[H]
    \centering
        %  \caption{ Effective mass plot $m^{\rm eff}_x=\frac{1}{a}\ln\frac{C_x(t)}{C_x(t+a)}$, as calculated from the  correlator $C_x(t)=\sum_{s_1,s_2} \langle x(t,s_1) x^*(0,s_2)\rangle$ of bosonic fields $x,x^*$ in  presence  of Wilson terms.} 
                \includegraphics[scale=0.65]{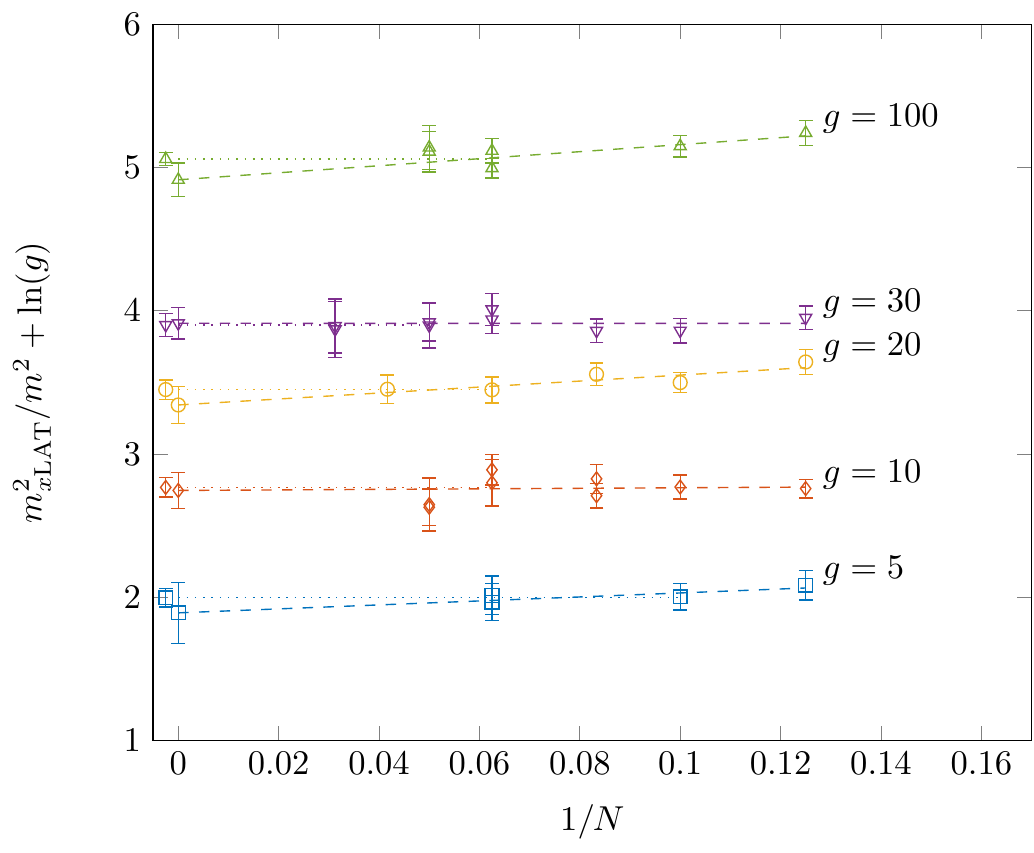}
\hspace{0.5cm}
              \hspace{0.5cm}
    \includegraphics[scale=0.65]{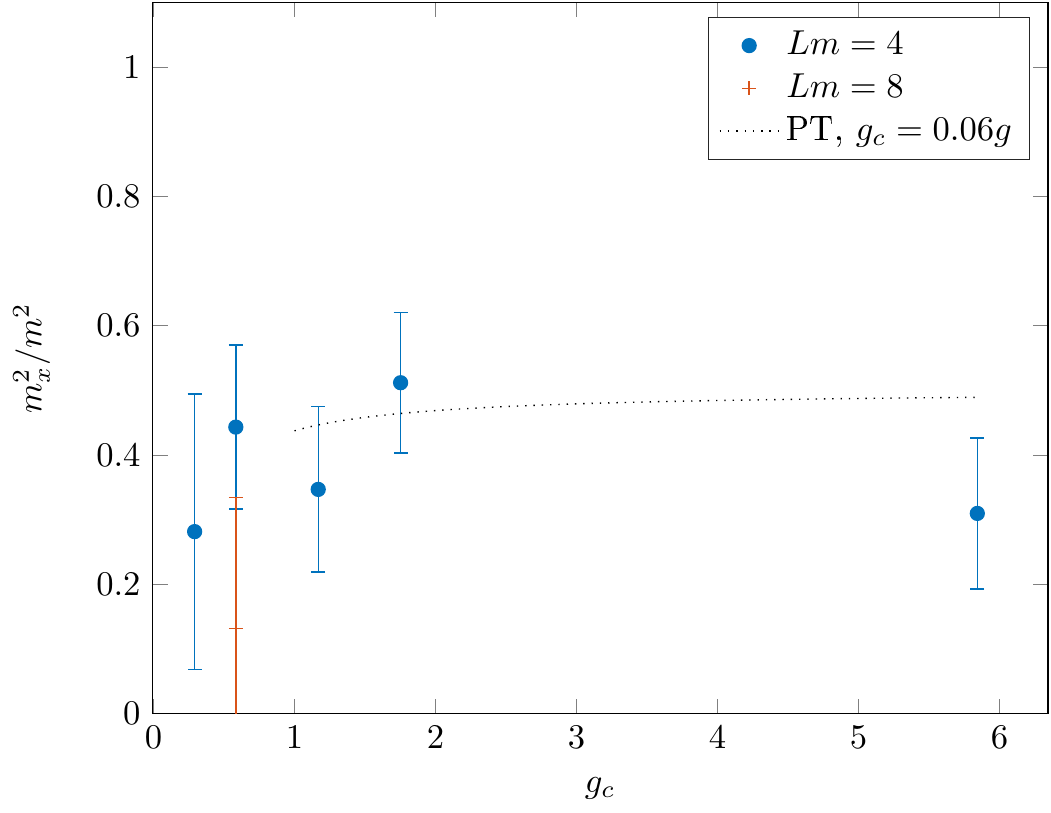}
      %  \caption{ Effective mass plot $m^{\rm eff}_x=\frac{1}{a}\ln\frac{C_x(t)}{C_x(t+a)}$, as calculated from the  correlator $C_x(t)=\sum_{s_1,s_2} \langle x(t,s_1) x^*(0,s_2)\rangle$ of bosonic fields $x,x^*$ in  presence  of Wilson terms.} 
                    \caption{
Plot of $m^2_{\rm x LAT}(N,g)/m^2 = m_x(g) + \mathcal{O}(1/N)$ and its continuum extrapolation, realized here using the $SO(6)$-breaking discretization \eqref{OFgenbreak}-\eqref{Wilsonshiftgenbreak}. Detailed explanation and comments as in Fig.~\ref{fig:correlator}.
                                     } 
            \label{fig:correlator_break}
\end{figure}

\begin{figure}[H]
    \centering
   \includegraphics[scale=1]{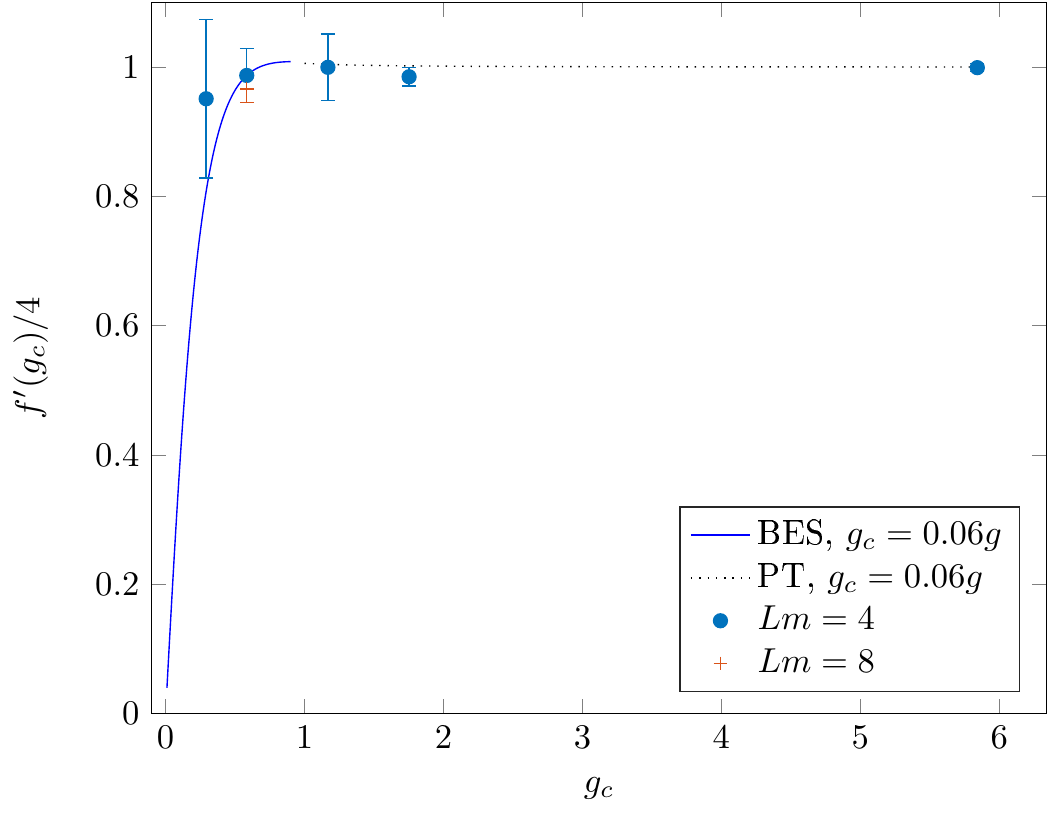}  
        \caption{Plot for $f'(g)/4$ as determined from the $N\to\infty$ extrapolation of \eqref{fit2}, realized here using the $SO(6)$-breaking discretization \eqref{OFgenbreak}-\eqref{Wilsonshiftgenbreak}. 
        Detailed explanation and comments as in Fig.\ref{fig:action_fin_g}.
 } 
     \label{fig:action_fin_g_break}
\end{figure}

 In this Appendix we collect results on simulations performed employing an alternative discretization, for which the fermionic operator reads
\begin{eqnarray}\nonumber
\!\!\!\!\!\!\!\!\!\!\!\!\!\!\!\!
\widetilde{O}_F\!\!=\!\!&\left(\begin{array}{cccc}
\!\!\!\!      \widetilde{W}_+      &\!\!\!\! - \mathring{p_0} \mathbb{1}      &\!\!\!\!  (\mathring{p_1}-i\frac{m}{2} )\rho^M \frac{z^M}{z^3}  &\!\!\!\! 0 \\
   \!\!\!\!   - \mathring{p_0}\mathbb{1}  &\!\!\!\! -  \widetilde{W}_+^{\dagger}          &\!\!\!\! 0        &\!\!\!\! \rho_M^\dagger(\mathring{p_1}-i\frac{ m}{2})\frac{z^M}{z^3} \\
   \!\!\!\!   -( \mathring{p_1}+i\frac{m}{2} )\rho^M \frac{z^M}{z^3}   &\!\!\!\! 0          &\!\!\!\! 2\frac{{z}^{M}}{{z}^{4}}\rho^{M}\left(\partial_{s}{x}-m\frac{{x}}{2}\right) +  \widetilde{W}_-    & \!\!\!\! - \mathring{p_0} \mathbb{1}-A^T\\
    \!\!\!\!  0      &\!\!\!\! -\rho_M^\dagger( \mathring{p_1}+i\,\frac{m}{2})\frac{z^M}{z^3}   &\!\!\!\! - \mathring{p_0}\mathbb{1}+A    & \!\!\!\! -2\frac{{z}^{M}}{{z}^{4}}\rho_{M}^{\dagger}\left(\partial_{s}{x}^*-m\frac{{x}}{2}^*\right) -  \widetilde{W}_-^\dagger
          \end{array}\right)\\
          \label{OFgenbreak}
          \end{eqnarray}
where the only change with respect to \eqref{OFgen} is in the Wilson term adopted, which now is
 \be\label{Wilsonshiftgenbreak}
\widetilde{W}_\pm = \frac{r}{2\,z^3}\,\big({\hat p}_0^2\pm i\,{\hat p}_1^2\big)\,\big(\rho_6\,z_m\,z^m+\rho_1\,(z^6)^2\big)\,,\qquad m=1,\cdots, 5~.
\ee
This discretization, which is the one employed in~\cite{POS2015} and for which simulation parameters are reported in Table \ref{t:runs2}, is consistent with lattice perturbation theory performed around vacua coinciding with one of six cartesian coordinates $u^M\,,~~M=1,\cdots,6$ (and no general linear combination of them). It also maintains all requirement listed in Section \ref{sec:discretization}, 
except for the one on $SO(6)$ invariance, which is explicitly broken (the other global symmetry of the model, $U(1)$ ,  is also broken). One can compare the continuum extrapolations of the two observables under investigation in the different discretizations, namely  Fig. \ref{fig:correlator} with  Fig. \ref{fig:correlator_break} for the $x$-mass and Fig. \ref{fig:action_fin_g} with  Fig. \ref{fig:action_fin_g_break} for the action. They agree within errors, which is strongly suggesting that the two discretizations lead to the same continuum limit. 

\begin{figure}[H]
   \centering
           \includegraphics[scale=0.65]{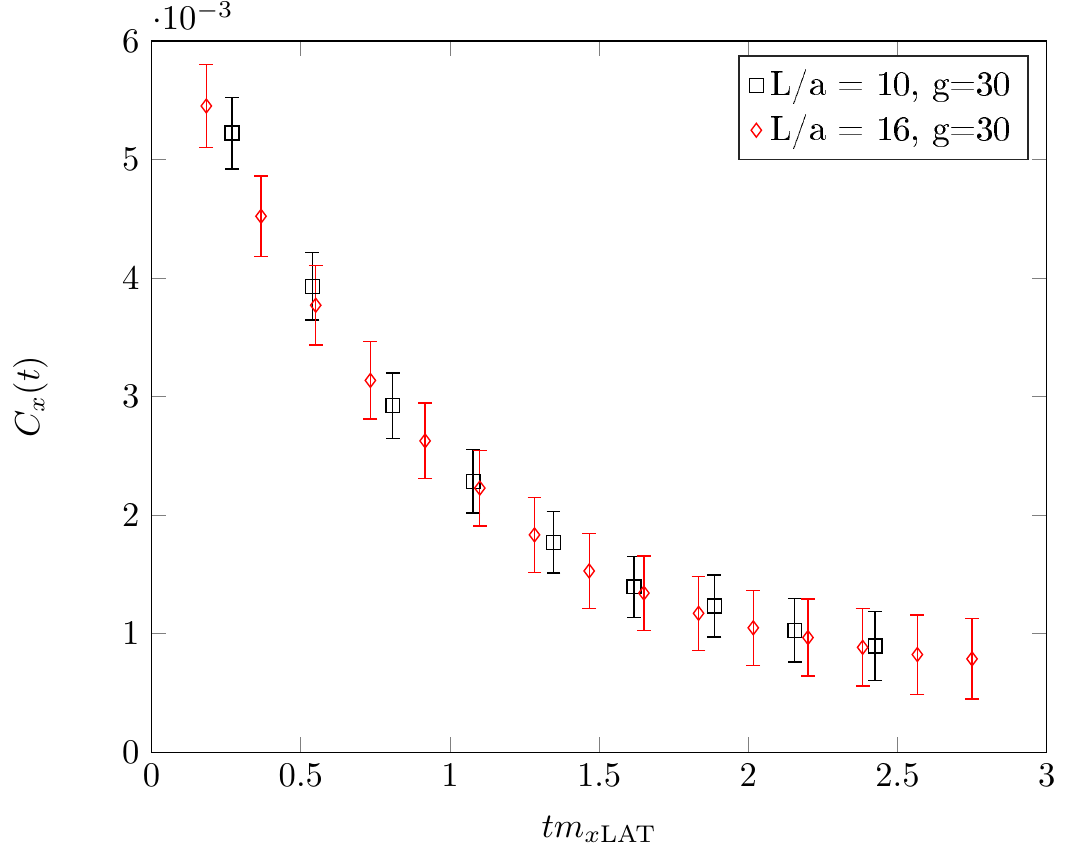}        
         \hspace{0.5cm}
         %  \caption{ } 
         \hspace{0.5cm}
          \includegraphics[scale=0.65]{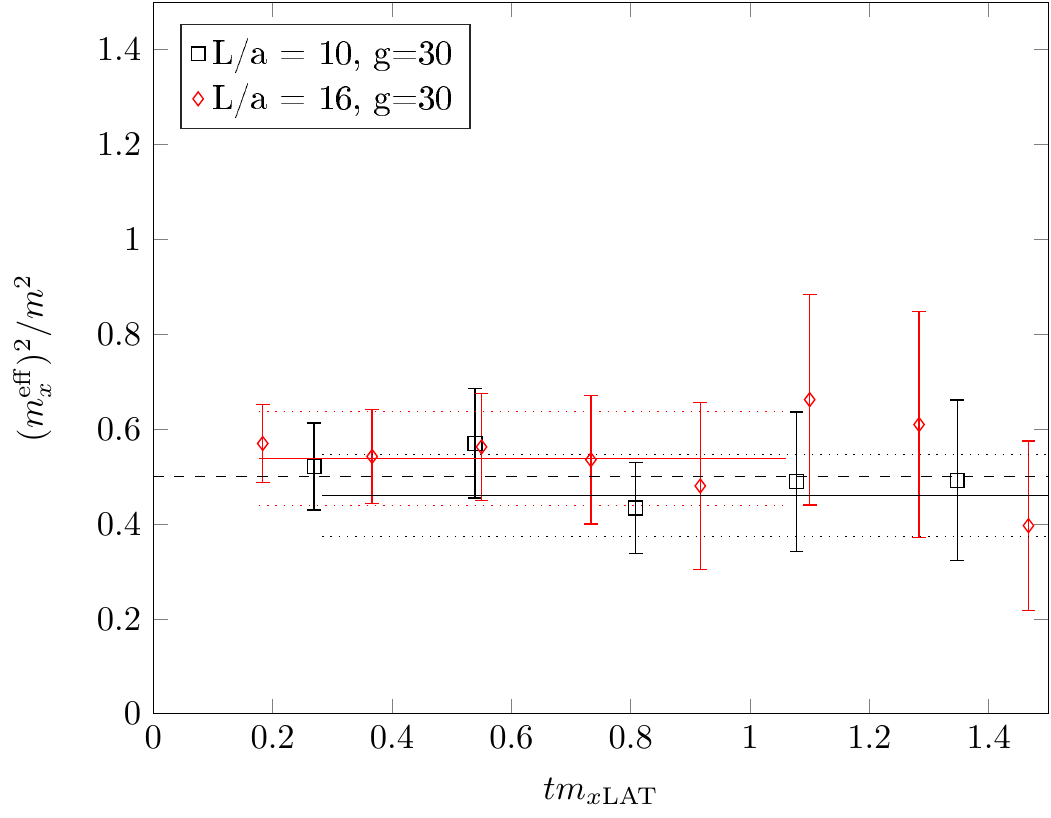}
           \hspace{0.5cm}
            \caption{Correlator and  mass for the $x$ field, realized here using the $SO(6)$-breaking discretization \eqref{OFgenbreak}-\eqref{Wilsonshiftgenbreak}. Detailed explanation and comments as in Fig.~\ref{fig:correlatormass}.} 
             \label{fig:correlatormass_break}
\end{figure}
\begin{figure}[H]
  \centering
     \includegraphics[scale=0.68]{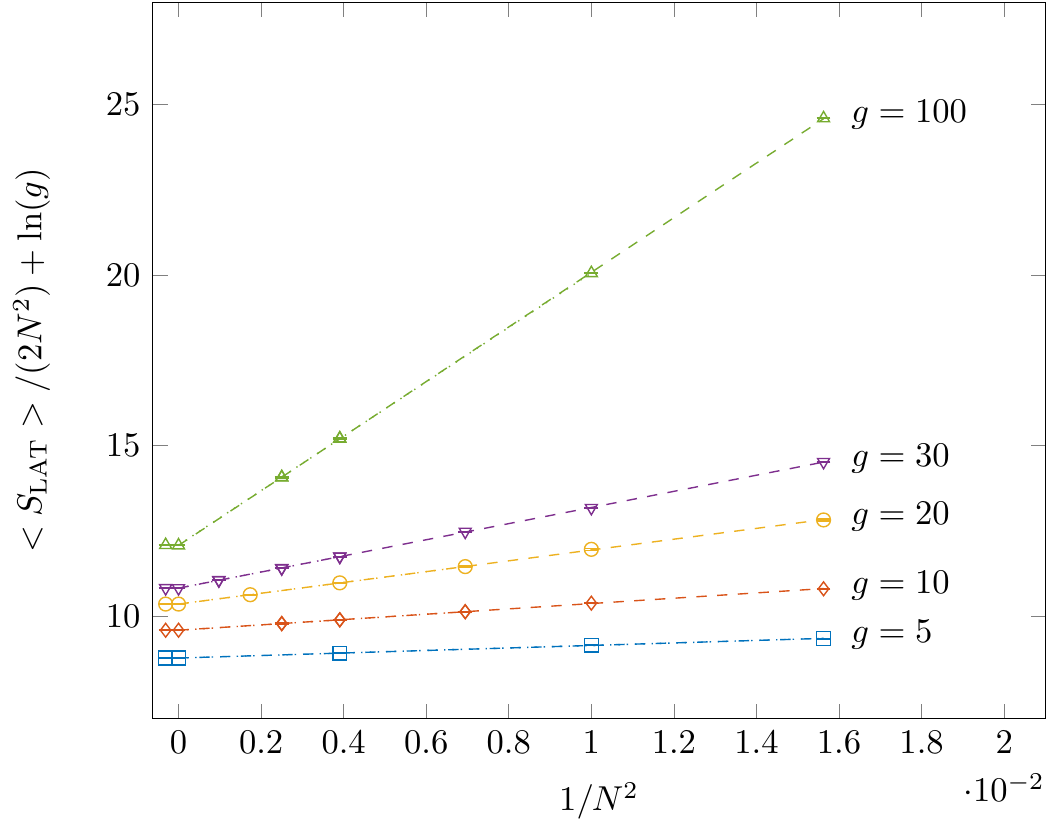}
      \hspace{0.5cm}
  \includegraphics[scale=0.71]{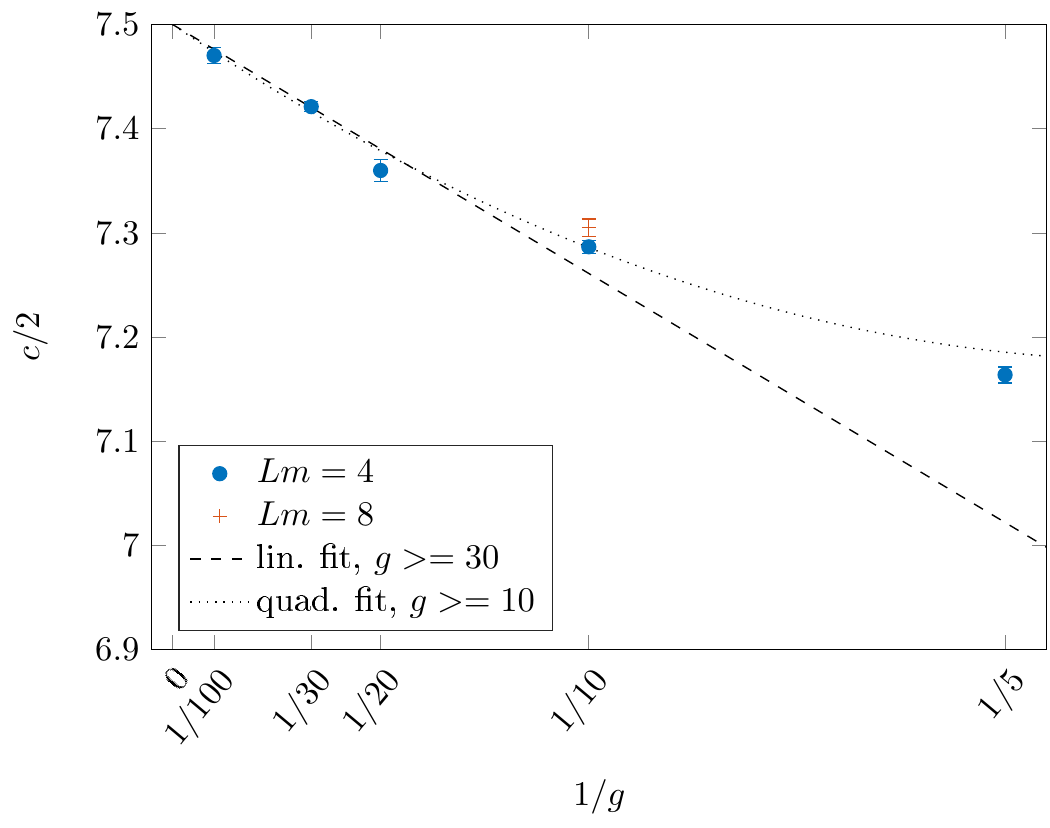}
        \caption{
Plots  of $\frac{\langle S_{\rm LAT}\rangle}{2 N^2}$ and its continuum extrapolation to determine $c/2$, realized here using the $SO(6)$-breaking discretization \eqref{OFgenbreak}-\eqref{Wilsonshiftgenbreak}. 
 Detailed explanation and comments as in Fig.~\ref{fig:constant}.
 }
     \label{fig:constant_break}
     \end{figure}
\begin{figure}[H]
  \centering
     \includegraphics[scale=1]{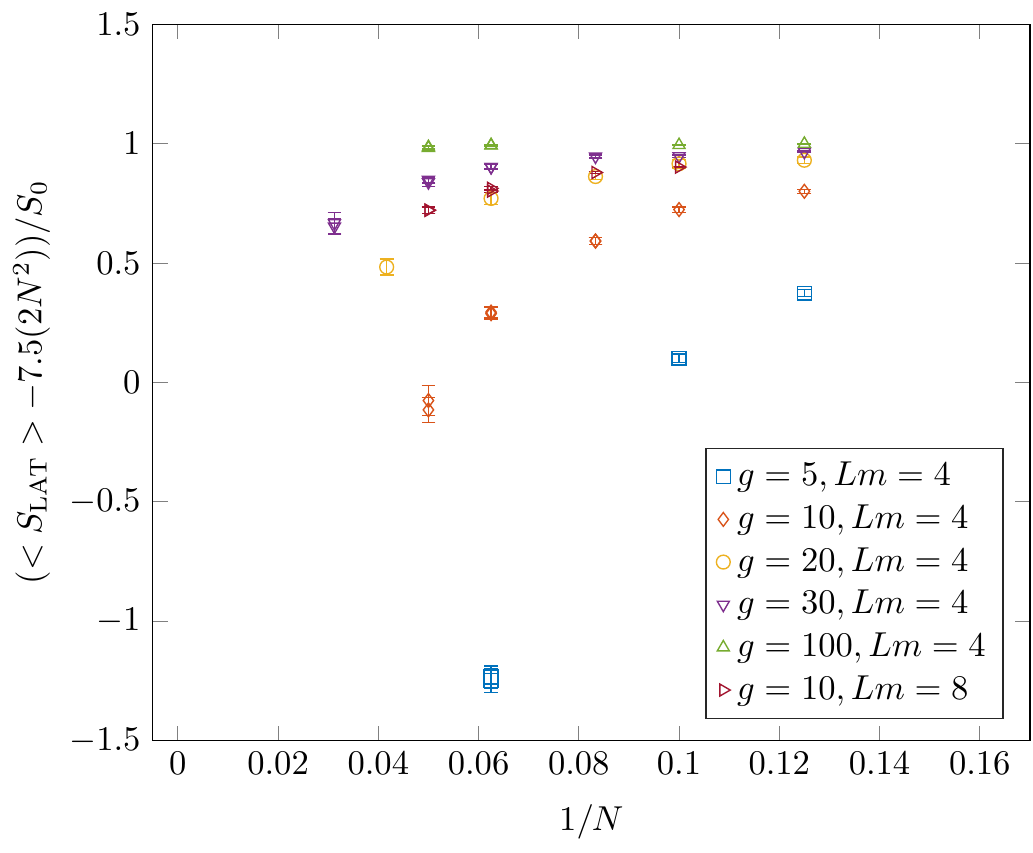}
        \caption{Plot of the ratio  $\frac{\langle S_{LAT}\rangle-\frac{c}{2}\, (2N^2)}{S_0} \equiv \frac{f'(g)}{4}$, realized here using the $SO(6)$-breaking discretization \eqref{OFgenbreak}-\eqref{Wilsonshiftgenbreak}. 
        Detailed explanation and comments as in Fig.~\ref{fig:action_div}.} 
     \label{fig:action_div_break}
\end{figure}
\begin{figure}[H]
  \centering
     \includegraphics[scale=1]{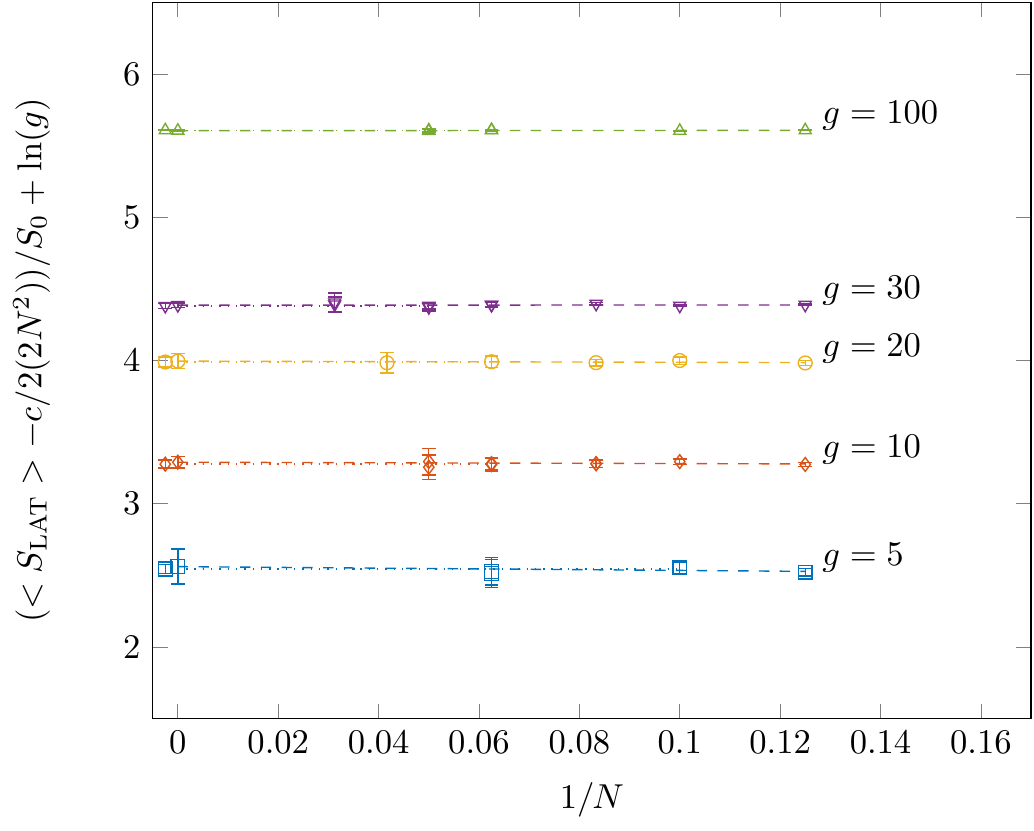}
          \caption{
Plots for the ratio $\frac{\langle S_{LAT}\rangle-\frac{c}{2}\, (2N^2)}{S_0}+\ln g $ as a function of $1/N$, realized here using the $SO(6)$-breaking discretization \eqref{OFgenbreak}-\eqref{Wilsonshiftgenbreak}. 
Detailed explanation and comments as in Fig.\ref{fig:action_fin_N2}.
           } 
     \label{fig:action_fin_N2_break}
\end{figure}
\begin{table}[H]
 \centering
\input{so6breaking/run_table.tex}
 \caption{
 Parameters of the simulations performed with the discretization \eqref{OFgenbreak}-\eqref{Wilsonshiftgenbreak}. The temporal extent $T$ is always twice the spatial extent, which helps studying the correlators.
 The size of the statistic after thermalization is given in terms of Molecular Dynanic Units (MDU) which equal an HMC trajectory of length one.
 The typical auto-correlation time of the correlators is given in the last column.
}
 \label{t:runs2}
\end{table}

% \section{Integral $\mathcal{I}(a)$}
% \label{app:integral}

\bibliographystyle{nb}
\bibliography{Ref_strings_lattice}

\end{document}

%% file: so6breaking/run_table.tex
\begin{tabular}{ccccccc}
\toprule
$g$ & $T/a\times L/a$ & $Lm$ & $am$ & $\tau_{\rm int}^{S}$ & $\tau_{\rm int}^{m_x}$ & statistic [MDU] \\
\midrule
  5 & $16 \times   8$ &  4  & 0.50000 & 0.8 & 2.7 & 900 \\
    & $20 \times  10$ &  4  & 0.40000 & 0.8 & 2.8 & 900 \\
    & $32 \times  16$ &  4  & 0.25000 & 2.0 & 8.1 & 950,950 \\
\midrule
 10 & $20 \times  10$ &  8  & 0.80000 & 1.1 & 2.2 & 900 \\
    & $24 \times  12$ &  8  & 0.66667 & 1.4 & 2.5 & 900 \\
    & $32 \times  16$ &  8  & 0.50000 & 2.4 & 5.8 & 750,750 \\
    & $40 \times  20$ &  8  & 0.40000 & 5.8 & 10.6 & 900,900 \\
    & $16 \times   8$ &  4  & 0.50000 & 0.8 & 1.9 & 900 \\
    & $20 \times  10$ &  4  & 0.40000 & 1.0 & 2.2 & 900 \\
    & $24 \times  12$ &  4  & 0.33333 & 1.1 & 2.6 & 900,900 \\
    & $32 \times  16$ &  4  & 0.25000 & 1.9 & 5.0 & 925,925 \\
    & $40 \times  20$ &  4  & 0.20000 & 7.8 & 11.7 & 925,925 \\
\midrule
 20 & $16 \times   8$ &  4  & 0.50000 & 8.7 & 2.7 & 1000 \\
    & $20 \times  10$ &  4  & 0.40000 & 10.9 & 2.3 & 1000 \\
    & $24 \times  12$ &  4  & 0.33333 & 4.7 & 2.0 & 1000 \\
    & $32 \times  16$ &  4  & 0.25000 & 6.5 & 3.3 & 850 \\
    & $48 \times  24$ &  4  & 0.16667 & 6.2 & 3.2 & 918 \\
\midrule
 30 & $16 \times   8$ &  4  & 0.50000 & 1.3 & 2.0 & 800 \\
    & $20 \times  10$ &  4  & 0.40000 & 1.2 & 2.1 & 800 \\
    & $24 \times  12$ &  4  & 0.33333 & 1.7 & 2.9 & 900 \\
    & $32 \times  16$ &  4  & 0.25000 & 2.7 & 4.1 & 950,950 \\
    & $40 \times  20$ &  4  & 0.20000 & 3.7 & 11.0 & 950,900 \\
    & $64 \times  32$ &  4  & 0.12500 & 6.9 & 31.1 & 579,900 \\
\midrule
100 & $16 \times   8$ &  4  & 0.50000 & 1.6 & 3.3 & 900 \\
    & $20 \times  10$ &  4  & 0.40000 & 2.0 & 3.8 & 750 \\
    & $32 \times  16$ &  4  & 0.25000 & 2.8 & 3.8 & 900,900 \\
    & $40 \times  20$ &  4  & 0.20000 & 6.2 & 10.4 & 900,900 \\
\bottomrule
\end{tabular}